\documentclass[letterpaper, aps, prd, twocolumn, superscriptaddress, nofootinbib, floatfix,showpacs]{revtex4}

\usepackage{graphicx}
\usepackage[svgnames,dvipsnames]{xcolor}
\usepackage{bm}
\usepackage{amssymb}
\usepackage{amsmath}
\usepackage{amsfonts}

\usepackage[utf8]{inputenc}

\graphicspath{{figures/},{./},{SingleStar/}}
\DeclareGraphicsExtensions{.pdf,.png,.jpg}

\newcommand{\newacronym}[3]{\newcommand{#1}[1]{#3##1 (#2)\renewcommand{#1}[1]{#2####1}}}
\newcommand{\newbareacronym}[3]{\newcommand{#1}[1]{#3 (#2)\renewcommand{#1}[1]{#2}}}

\newcommand{\sun}{{\mathord\odot}}
\newcommand{\scriplus}{\ensuremath{\mathcal{I}^{\mathord +}}}

\newcommand{\dint}{\mathop{}\!\mathrm{d}}
\newcommand{\sfrac}[2]{#1/#2}

\newcommand{\gridspacing}{\Delta x}
\newcommand{\enth}{h^{e}}

\newcommand{\code}[1]{{\tt #1}}

\newacronym{\SpEC}{\code{SpEC}}{Spectral Einstein Code}
\newcommand{\bam}{\code{BAM}}
\newacronym{\ligo}{LIGO}{Laser Interferometer Gravitational-Wave
Observatory}
\newcommand{\virgo}{Virgo}
\newacronym{\cfl}{CFL}{Courant-Friedrichs-Lewy}
\newacronym{\kagra}{KAGRA}{Kamioka Gravitational Wave Detector}
\newacronym{\tov}{TOV}{Tolman-Oppenheimer-Volkoff}
\newacronym{\eos}{EOS}{equation of state}
\newacronym{\weno}{WENO}{weighted essentially non-oscillatory}
\newacronym{\nsns}{NSNS}{neutron star -- neutron star}
\newbareacronym{\bhns}{BHNS}{black hole -- neutron star}
\newacronym{\bbh}{BBH}{binary black hole}
\newacronym{\ns}{NS}{neutron star}
\newacronym{\bh}{BH}{black hole}
\newacronym{\eob}{EOB}{effective one body}
\newacronym{\cce}{CCE}{Cauchy-characteristic extraction}
\newacronym{\amr}{AMR}{adaptive mesh refinement}
\newacronym{\adm}{ADM}{Arnowitt-Deser-Misner}
\newacronym{\ah}{AH}{apparent horizon}
\newacronym{\hmns}{HMNS}{hypermassive neutron star}
\newacronym{\PN}{PN}{post-Newtonian}
\newacronym{\gw}{GW}{gravitational wave}
\newacronym{\gr}{GR}{General Relativity}
\newacronym{\hll}{HLL}{Harten-Lax-van Leer}
\newcommand{\Lev}[1]{\texttt{Lev#1}}
\newcommand{\conj}[1]{{#1}^{\star}}
\newcommand{\mpct}[1]{\,\text{#1}}

\newcommand{\igrid}{i}
\newcommand{\rgrid}{r}
\newcommand{\thetagrid}{\theta}
\newcommand{\phigrid}{\phi}

\newcommand{\TAPIRaddr}{TAPIR, Walter Burke Institute for Theoretical
  Physics, MC 350-17, California Institute of Technology, 1200 E
  California Blvd., Pasadena California 91125, USA}
\newcommand{\AEIaddr}{%
  Max Planck Institute for Gravitational Physics
  (Albert Einstein Institute),
  Am M\"uhlenberg 1, Potsdam-Golm, 14476, Germany}
\newcommand{\JPLaddr}{%
  Jet Propulsion Laboratory, California Institute of Technology, 4800
Oak Grove Dr., Pasadena California, 91109, USA}
\newcommand{\cornelladdr}{%
  Cornell Center for Astrophysics and Planetary Science, Cornell University,
Ithaca, New York 14853, USA}
\newcommand{\citaaddr}{%
  Canadian Institute for Theoretical
Astrophysics, University of Toronto, Toronto, Ontario M5S 3H8, Canada}

\begin{document}

\title{Simulations of inspiraling and merging double neutron stars
using the Spectral Einstein Code}
 
\author{Roland Haas}
\affiliation{\AEIaddr}
\affiliation{\TAPIRaddr}

\author{Christian D.\ Ott}
\affiliation{Yukawa Institute for Theoretical Physics, Kyoto University, Kyoto, Japan}
\affiliation{\TAPIRaddr}

\author{Bela Szilagyi}
\affiliation{\JPLaddr}
\affiliation{\TAPIRaddr}

\author{Jeffrey D.\ Kaplan}
\affiliation{\TAPIRaddr}

\author{Jonas Lippuner}
\affiliation{\TAPIRaddr}

\author{Mark A.\ Scheel}
\affiliation{\TAPIRaddr}

\author{Kevin Barkett}
\affiliation{\TAPIRaddr}

\author{Curran D.\ Muhlberger}
\affiliation{\cornelladdr}

\author{Tim Dietrich}
\affiliation{\AEIaddr}

\author{Matthew D. Duez}
\affiliation{Department of Physics \& Astronomy,
Washington State University, Pullman, Washington 99164, USA}

\author{Francois Foucart}
\affiliation{\citaaddr}
\affiliation{Lawrence Berkeley National Laboratory, 1 Cyclotron Rd,
Berkeley, California 94720, USA}

\author{Harald P. Pfeiffer}
\affiliation{\citaaddr}

\author{Lawrence E. Kidder}
\affiliation{\cornelladdr}

\author{Saul A. Teukolsky}
\affiliation{\cornelladdr}

\date{\today}

\begin{abstract}
We present results on the inspiral, merger, and postmerger evolution of a
\nsns{} system.
Our results are obtained using the hybrid pseudospectral finite volume
\SpEC{}.
To test our numerical methods, we evolve an equal-mass system
for $\approx 22$ orbits before merger. This waveform is the longest waveform
obtained
from
fully general-relativistic simulations for \nsns{s} to date.
Such long (and accurate) numerical waveforms are required to further improve
semianalytical models used in gravitational wave data analysis, for example
the effective one body models.
We discuss in detail the improvements to \SpEC{'s} ability
to simulate \nsns{} mergers, in particular
mesh refined grids to better resolve the merger and postmerger phases.
We provide a set of consistency checks and compare our results
to \nsns{} merger simulations with
the independent \bam{} code.  We find agreement between them, which
increases confidence in results
obtained with either code.
This work paves the way for future studies
using long waveforms and more complex microphysical descriptions of neutron
star matter in \SpEC{}.
\end{abstract}


\maketitle

\section{Introduction}
\label{section:introduction}

The Advanced \ligo{} demonstrated
its capability of measuring \gw{} signals coming from compact binary
systems and opened up a new window for astrophysical observations.
Although the first detected \gw{}
signal~\cite{LIGOVirgo2016a,TheLIGOScientific:2016qqj} was emitted by
a binary made up of two \bh{s}, \nsns{} systems are
promising sources~\cite{Abadie:2010cf}.  With further upgrades of
\ligo{} and with the entire \gw{} network consisting of \ligo{},
\virgo{}, and the \kagra{}~\cite{Anderson:2007km, Dominik:2014yma}
operating, between $0.2$ and $200$ \nsns{} mergers per
year~\cite{Abbott:2016wya} are expected to
be observed.

The \gw{s} emitted during the inspiral and merger contain unique
information about the binary's properties and
about each binary
constituent. In the case of NS systems,
information about the \eos{} at supranuclear densities can be
obtained that is not easily obtainable otherwise~\cite{Anderson:2007km}.

The \gw{} signal of a \nsns{} coalescence can be roughly separated into
three phases: the \emph{inspiral} phase in which the \ns{s} approach
each other due to the emission of \gw{s}, the \emph{merger} phase in
which the stars come in contact and form a single object,\footnote{We
  define the actual moment of merger as the time at which the \gw{}
  strain has its maximum.}  and the \emph{postmerger} phase in which
the remnant can, depending on its mass and the \eos{} of the star,
 (i) collapse promptly to a \bh{} leading to a
characteristic ringdown \gw{} signal; (ii) form a \hmns{} which is
stabilized primarily by angular momentum over secular time scales
before \bh{} formation; (iii) form a long-term stable
(supramassive or massive) \ns{} if the total mass of the system is
sufficiently low. See,
eg.~\cite{Baumgarte:1999cq,Hotokezaka:2011dh,Takami:2015,Bernuzzi2015} for
studies of the waveform spectra and classification of the outcomes.
The inspiral part of the signal sweeps
  through the most sensitive band of current \gw{} detectors.
The merger, postmerger, and ringdown parts of the
  signal are at high frequencies and are difficult to observe unless
  the event is very close.

For most of the lifetime of the binary, the \ns{s} are well separated
and the signal is almost sinusoidal with slowly changing frequency.
During this phase, the wave signal can be approximated by
\PN{} theory to very high accuracy (see~\cite{Blanchet2014}
and references therein).
The binary orbit evolves adiabatically under the influence of the
radiation reaction force.  Close to merger and in the postmerger
phase, \PN{} theory is no longer valid and numerical
relativity simulations are needed to correctly capture the fast dynamics and
construct \gw{} templates, see,
e.g.,~\cite{Takami:2015,Clark:2015zxa,Bauswein:2015vxa,Bernuzzi2012b,Bernuzzi2015}.

This late, high-frequency part of the \gw{} signal is most interesting since it
is directly affected
by the star's \eos{}.
The \eos{}, via the tidal deformability of the \ns{s}, leaves a clear
imprint on the
late inspiral and early merger phases of the \gw{} signal~\cite{Read2009b,Hinderer2010,Markakis:2010mp,ReadEtAl2013,
Maselli:2013rza}. In addition, also the postmerger frequency and the
time evolution of the \gw{} signal from the merger remnant can constrain the
\eos{} via \gw{}
observations, see, e.g., \cite{Takami:2014zpa,Bauswein:2015vxa,Bernuzzi2015,
Kastaun:2014fna,DallOsso:2014zza}.

While the original \PN{} expansion breaks down before merger,
the \eob{} model~\cite{Buonanno99,Buonanno00,DJS00,Damour01c} provides
techniques to extend the range of applicability of \PN{} theory into the
late inspiral phase, also including tidal effects~\cite{Bernuzzi2015,Hinderer:2016a}.
The tidal parameters of the model can be
linked to the parameters of the \eos{} of the \ns{s}, making
it possible to determine \eos{} parameters from the \gw{}
signal. Similarly, there is interest in ``universal'' relations
between observable quantities that are independent of the
\eos{}~\cite{Yagi:2013bca,ReadEtAl2013,Chatziioannou:2015uea,Doneva:2015hsa,Bernuzzi:2014kca,Bernuzzi2015}
and the breakdown of this universality~\cite{Haskell:2013vha}.
Models accurately describing \ns{} coalescences beyond the merger
are still missing and only numerical simulations in full general relativity
can give reliable information
about this stage, but see~\cite{Clark:2015zxa} for a first attempt of a
reduced-order
model of the postmerger waveform.

Numerical simulations are needed to validate and calibrate \eob{}
models, e.g.~\cite{Baiotti2011, Bernuzzi2012b}. This is possible only
if (i) numerical waveforms have a sufficient length and span multiple
orbits, so that \PN{} approximations are valid at the
beginning of the simulated period, and (ii) they are sufficiently accurate,
i.e., having small eccentricities and small phase errors. Several
authors~\cite{Bernuzzi2012b,hotokezaka:12,ReadEtAl2013,
  Favata:2013rwa,Wade:2014vqa} have studied the detectability of tidal
effects in detail, investigating the errors and uncertainties and the
effect of different equations of state on the wave signal.  %
The results of these studies underscore the importance of a careful
assessment of numerical errors and the influence of the numerical scheme
used on the gravitational waveform. The effect of a lack of resolution in
particular can mimic
physical effects
such as the effect of tidal interactions which primarily manifests as an
increased rate of inspiral of the binary.
Some of us recently presented the
results for spinning \nsns{} inspirals~\cite{Tacik:2015tja} and merger
simulations of \ns{} binaries including neutrino
transport~\cite{Foucart:2015} using the \SpEC{}.  \SpEC{} simulations
involving
\bhns{} were
performed~\cite{Duez:2008rb,Duez:2009yy,FoucartEtAl:2011,Deaton2013,Foucart:2014nda}
and
\bbh{}~\cite{Scheel2009,Lovelace:2010ne,MacDonald:2012mp,Mroue:2013PRL,Blackman:2015pia,Ossokine:2015vda}
simulations
using \SpEC{} have a long history.  This paper follows the line of
work focusing on the accuracy and feasibility of constructing
sufficiently long and accurate \gw{} templates.  For this purpose, we
extended \SpEC{'s} \nsns{s} simulating capabilities.  \SpEC{} employs
a hybrid approach using pseudospectral methods for the spacetime
evolution and finite volume or finite differencing methods for the
hydrodynamical variables.  This allows us to achieve very high phase
accuracy at low computational costs for the spacetime part of the
evolution and to exploit well-tested, stable high-resolution shock
capturing methods for the fluid variables.  \SpEC{} uses a comoving
coordinate system which reduces movement of the \ns{s} on the grid and
therefore reduces possible errors accumulated in moving-box
mesh-refined schemes by the restriction and prolongation operation as
well as Eulerian advection errors.  This paper provides numerical
tests of these methods in \SpEC{,} paving the way for a more
systematic comparison with existing codes to simulate
\nsns{}~\cite{Giacomazzo2007,YamamotoShibata2008,Thierfelder:2011yi,Radice:2013cba,moesta:14a}
systems and long \nsns{} inspiral simulations involving more realistic
\eos{}.  A first step toward such a comparison is made by comparing
our data with a \bam{} waveform
of~\cite{Bernuzzi:2014owa}. We find
very good agreement and a phase differences below $0.25\,\mathrm{rad}$
up to the end of the inspiral phase.

As outlined in the paragraph
above, not only is sufficient accuracy  needed
in \nsns{} simulations, the waveforms also need to span multiple
orbits before merger to be useful for semianalytical waveform
modeling.  Here we consider an equal mass binary system with an
initial coordinate separation
of $81\,\mathrm{km}$ and baryon mass of $M_0 = 1.779\,M_\sun$ of each star,
which results in more than 22 orbits before
merger, i.e., 44~\gw{} cycles before merger.  This is to date the
longest \nsns{} merger simulation and the resulting waveform has
already been used for the analysis of~\cite{Barkett2015}.  Such long
simulations can be achieved due to the small computational expense for
evolving the metric variables with our pseudospectral approach as well
as the small fluid grids which cover the regions around the \ns{s} only,
instead of the whole simulation volume.

This paper is organized as follows. Section~\ref{sec:methods} presents the
methods used to evolve the spacetime and hydrodynamics sectors of the
simulations. Section~\ref{sec:initial_data} describes how we construct initial
data for \nsns{} systems. Section~\ref{sec:results} presents results on
the convergence and diagnostics on the quality of the computed waveforms.
We conclude in Section~\ref{sec:conclusions}.
In Appendixes \ref{sec:appendix_TOV} and \ref{sec:appendix_collapse},
we
present convergence tests for a single \tov{} star and
investigate the collapse dynamics of isolated \ns{s}, respectively.

Unless stated otherwise, all results use $G = c = 1$ and masses are given
in multiples of the solar mass $M_\sun$. $\nabla_\alpha$ is used to denote
the covariant derivative compatible with the 4-metric
$g^{(4)}_{\alpha\beta}$ and we use the signature convention of~\cite{Wald84}.

\section{Methods}
\label{sec:methods}

\subsection{Two-domain approach to general-relativistic hydrodynamics}
\label{sec:two_domain_approach_to_general_relativistic_hydrodynamics}

In \SpEC{} we  use a mixed approach to solve Einstein's equations in the
generalized harmonic formulation coupled to
matter~\cite{Duez:2008rb,FoucartEtAl:2011}.
We solve the evolution equations for the spacetime metric $g^{(4)}_{\alpha\beta}$
using spectral methods as described
in~\cite{Buchman:2012dw,Lovelace:2011nu,Scheel2009,Kidder2000a,Lindblom2006,
Scheel2006,Szilagyi:2009qz,Lovelace:2010ne,Hemberger:2012jz,Ossokine:2013zga}
while the fluid equations are solved using high-resolution shock-capturing
methods described in~\cite{Duez:2008rb,FoucartEtAl:2011,Foucart:2013a}. 
The \ns{} material is modeled as a perfect fluid with rest mass density
$\rho_0$, pressure $P$, specific internal energy $\epsilon$, and 4-velocity
$u^\alpha$, so that the stress-energy tensor is given by
\begin{align}
T_{\alpha\beta} = \rho_0 \enth u_\alpha u_\beta + P g^{(4)}_{\alpha\beta}
\mpct{,}
\end{align}
where $\enth = 1 + \epsilon + P/\rho_0$ is the relativistic specific enthalpy.
The evolution equations for the conserved hydrodynamical
variables $D = \sqrt{-g^{(4)}} u^t \rho_0$, $\tau = E - D$,
$S_k = \sqrt{-g^{(4)}} {T^t}_k$ follow from
conservation of stress-energy $\nabla_\alpha T^{\alpha\beta} = 0$ and
conservation of baryon number $\nabla_\alpha \left( D u^\alpha \right) = 0$, 
where $g^{(4)}$ is the determinant of the $4$-metric and
$E = \sqrt{-g^{(4)}} T^{tt}$.
In this paper, we split the pressure and specific internal energy in cold and
thermal pieces due to cold (nuclear force) and thermal contributions,
respectively,
\begin{align}
\epsilon &= \epsilon_{\text{cold}}(\rho_0) + \epsilon_{\text{thermal}}\mpct{,} \\
P &= P_{\text{cold}}(\rho_0) + \left(\Gamma-1\right)
      \rho_0\epsilon_{\text{thermal}}\mpct{,} \\
P_{\text{cold}} &= \kappa \rho_0^\Gamma\mpct{,} \\
\epsilon_{\text{cold}} &= P_{\text{cold}} / [ \rho_0\,(\Gamma-1) ]\mpct{,}
\end{align}
where $\kappa$ and $\Gamma$ are the polytropic constant and the adiabatic
index, respectively. In this paper we choose $\Gamma=2$, $\kappa =
123.6 M_\sun^2$ which \eos{} can support a nonrotating \ns{} of baryonic mass up to
$2.0\,M_\sun$. Note, however,
that SpEC can handle more general \eos{} than the one presented
here~\cite{Duez:2009yy,Deaton2013,Muhlberger2014,Foucart:2014nda,FoucartM1:2015}.
We use the fifth-order \weno{} reconstruction
method of~\cite{Liu1994200,Jiang1996202,Muhlberger2014,WENO5Comment}, a
\hll{} Riemann solver~\cite{HLL} to compute numerical fluxes at cell
interfaces, and a two-dimensional nonlinear root finding algorithm~\cite{Muhlberger2014} to
recover the primitive variables from the conserved variables at the
beginning of each time step.

Time
integration is performed using the method of lines~\cite{Hyman:1976cm} and
a third order Runge-Kutta (RK3) method with adaptive
step size control
based on errors in both spacetime and
hydrodynamical variables estimated by comparing second and third order
accurate time stepper updates. The metric evolution couples to the
hydrodynamical
evolution via the stress-energy tensor $T_{\mu\nu}$ and the hydrodynamical
evolution equations directly involve the metric and its first derivatives.
We interpolate spacetime and hydrodynamical variables between
spectral and finite volume grids at the end of each full time step as
described in detail in~\cite{Duez:2008rb} using almost-spectral
interpolation~\cite{Boyd:1992fac} to interpolate from the spectral grid to
the finite volume grid and monotonicity preserving polynomial interpolation
to interpolate between finite volume and spectral grids.
Values for
intermediate Runge-Kutta substeps are obtained via extrapolation in
time.

Since~\cite{Foucart:2013a} was published, we optimized this
communication scheme to reduce the amount of data sent between compute
nodes resulting in improved speed and scalability of the code.  These
changes include a reduction in the number of times \SpEC{'s} internal
dependency tracking recomputes quantities, limits copying of data in
memory, and significantly speeds up the transformation between
collocation point and spectral coefficient representations of
data. \SpEC{} now
interleaves communication and computation when
interpolating data between the spectral and the finite volume grids
after each time step. The
  number of explicit \code{MPI}
barriers has also been reduced. Beyond these infrastructure
changes, the basic setup described in~\cite{Duez:2008rb,Foucart:2013a}
remains the same.

\subsection{Comoving coordinate system}
\label{sec:comoving_coordinate_system}
\SpEC{} uses a dual frame method~\cite{Scheel2006,Hemberger:2012jz} to solve
Einstein's equations and the fluid equations. It uses explicit coordinate
transformations to map between a set of inertial (physical) coordinates in
which the \ns{s} orbit and approach each other and a
set of grid coordinates in which the \ns{s} remain at a fixed coordinate
location. Once a \bh{} is formed the coordinate transformation is also used
to map the excision surface inside the \ah{} to an excision sphere of
constant radius in grid coordinates.

The finite volume grid is linked to the spectral grid via a final,
piecewise constant in time coordinate transformation.
During the inspiral phase of the simulation, the \ns{s} are separated
by a vacuum region that gradually shrinks as the stars approach each
other and that does not need to be evolved with the matter evolution
equations. See Sec.~\ref{sec:numerical_setup} for a detailed
description of the grid setup during the individual phases of the
simulation. A single finite volume grid that covers both \ns{s} as
used in the \bh{}-\ns{} simulations in~\cite{FoucartEtAl:2011}
would thus be inefficient. Instead we cover each star with a separate
cubical finite volume grid. In this paper, we choose the boxes to be
initially $1.25$ times the diameter of the \ns{}. We track the motion
of each star and follow the inspiraling stars with the grids by
adjusting the grid locations.  For the purpose of tracking the stars,
we define each star's position to be the centroids of the rest mass
distribution,
\begin{align}
X^i_{\text{CM}} &= \int x^i\,D\,\dint^3 x
\mpct{,}
\end{align}
in each of the disjoint finite volume grids. Since the grid patches follow
the stars and all fluxes are expressed in the frame comoving with the grid,
the fluid velocities are small, which improves the accuracy of Eulerian
finite volume
codes~\cite{Tchekhovskoy:2007zn,Springel:2009aa,Wadsley:2008aa}.

We employ the remapping criteria outlined in~\cite{Foucart:2010eq} to
control the volume covered by the grid. As the \ns{s} spiral inwards, we
map the center of each \ns{} from its current position in the
physical (inertial) frame to a fixed location in the grid frame in which the
spectral basis functions are defined. Since the physical separation between
the \ns{s} shrinks with time, but their separation in the grid frame stays
constant, the \ns{s} appear to grow (cover more grid points) in the grid
frame. This gradually brings the surface of the \ns{s} closer to the grid
boundaries and eventually the outer layers of the \ns{s} reach the grid
boundaries and matter flows off the grid.

We measure the flux of matter,
\begin{align}
\dot M_0^{(i)} = \int_{S_{i, A}} D\,\frac{u^i}{u^t}\,\dint^2 x
\mpct{,}
\end{align}
through each of the outer boundaries $S_{i,\text{outer}}$ of the
finite volume grid and through a set of interior surfaces
$S_{i,\text{inner}}$, located approximately $44\%$ of the distance from
the center to the outer boundary. If too much ($\dot M_0^{(i)} >
2\times10^{-8}$) matter flows out of the grid through the
$i$-direction outer surface $S_{i,\text{outer}}$, we expand the grid
along the $i$-direction; if not enough matter ($|\dot M_0^{(i)}| <
3\times10^{-7}$, where $|\dot M_0^{(i)}|$ counts the total amount of
matter passing through the surface ignoring direction) flows through
the $i$-direction inner surface $S_{i,\text{inner}}$, we contract the
grid. We then interpolate the evolved variables onto the new
grid. During the inspiral this procedure typically keeps the outer
layers of the \ns{} with a rest mass density $\gtrsim 10^{-4}$ of the
density in the center of the star $\rho_{0,\text{central}}$ inside of
the grid, while lower density material may flow off the grid.  In
summary, this remapping procedure ensures that the total amount of
matter leaving the domain is controlled and that the grid stays as
small as possible.  A side effect of this procedure is that the
effective resolution changes during the simulation.  When we expand or
shrink the grid in response to the remapping criteria, the number of
grid points is kept fixed but the physical volume covered by the grid
changes, which leads to a discrete jump in the effective
resolution. In our simulation the typical jump in resolution due to grid
changes is approximately $10\,\%$ which can be seen in the discrete jumps in
Fig.~\ref{fig:dx}.
In
addition to these discrete jumps the inspiral of the \ns{s} toward
each other causes a continuous increase of resolution since the
physical area covered by the fixed number of grid points shrinks as
the \ns{s} approach each other in the inertial frame.

Finally, while the interpolation algorithm used is not strictly mass
conservative, the
remapping happens infrequently enough and with small enough incremental change
in the grid size that the effect on the total rest mass is $< 10^{-5}$ of
the total rest mass of the system over the course of the simulation. This is
several orders of magnitude lower than the amount of material ($\approx
10^{-3} M_\sun$) lost through the outer grid boundaries until an apparent
horizon forms, see Fig.~\ref{fig:rest-mass-conservation}. After horizon
formation any matter outside of the horizon
accretes rapidly onto the \bh{}.

\subsection{Mesh refinement}
\label{sec:mesh_refinement}
\SpEC{} employs adaptive mesh refinement in the spectral sector of the
evolution equations, adjusting both the order of the spectral basis
functions used as well as splitting subdomains into smaller subdomains
as required to achieve a desired truncation error. Spectral \amr{}
is described
in~\cite{Szilagyi:2014fna} to which we refer the reader for details.

In the finite volume sector,
in order to resolve both the region around each \ns{}
and cover a large enough volume
to capture outflows and the remnant disk that forms after merger of the
\ns{s} and \bh{}
formation, we employ a variant of the mesh
refinement techniques commonly used in compact binary merger
simulations~\cite{Foucart:2014nda}.
Often mesh refinement~\cite{Schnetter2003b,Fryxell:2000zz,Anderson:2006ay,YamamotoShibata2008}
is used not only to increase the resolution in
regions of interest but also to move the region of higher resolution along
with the object.
In this approach, as the \ns{s} move through the grid and get close to the
current
edge of the high resolution grid, new grid points are created and populated
with data interpolated from the coarse grid in front of the \ns{} and
no longer needed points are destroyed once the \ns{} has
passed.
Such an interpolation step necessarily leads to a loss of accuracy and great
care needs to be taken to preserve physically conserved quantities such as
rest mass as well as---in the absence of general relativity---energy and
momentum~\cite{Berger:1989,Baiotti:2010ka,Etienne:2010ui,East:2011aa,reisswig:13a,Dietrich:2015iva}.

In \SpEC{}, on the other hand, because of its
comoving coordinate system, the \ns{s} are
stationary on the grid during the inspiral phase and no mesh motion is
required. During most of the inspiral we use only a single resolution
in the grid patches that surround each \ns{} and no mesh refinement.
Eventually, however, the \ns{s} approach each other close enough such
that their individual grid patches overlap.  Rather than continue
evolving in the presence of overlapping grids as,
e.g.\ in~\cite{YamamotoShibata2008}, we create a single refined grid
hierarchy that contains both \ns{s}.  Since we continue to track the
rotation of the \ns{s} around each other but stop tracking their
separation, the \ns{s} appear to move directly toward each other on
the grid, and we again avoid having to create and destroy grid points
to follow the \ns{s} with a high resolution grid patch.

Our current implementation of mesh refinement in \SpEC{} uses
vertex-centered grid points such that for a factor of 2 difference in
resolution on
coarse and fine grids, every second fine grid point coincides with a coarse
grid point.  For the current set of simulations we employ only two
refinement levels. The code, however, is not restricted to this and supports
an arbitrary number of refinement levels in an arbitrary number of grid
patches. We currently do not employ subcycling in time: all refinement
levels step forward in time with the same time step size. This is similar to,
e.g.~\cite{MacNeice:2000aa}, but differs from the approach
in~\cite{Schnetter2003b}. In choosing to not implement subcycling in time
we sacrifice efficiency of the simulation for code simplicity. In
the current approach, we are able to leverage the existing multidomain
infrastructure in \SpEC{}
to implement the three required data
movement operations~\cite{Berger1984}: ``synchronization,'' ``restriction,''
and ``prolongation'' at grid boundaries as well as where fine and coarse grids
overlap. In \SpEC{}, synchronization is the exchange of grid point data
between grid patches that make up a single refinement level. It provides
data in neighboring grid cells required for \weno{} reconstruction of the
variables to cell interfaces. It never moves data between different
refinement levels, and since all grid patches are aligned, synchronization
is a
straightforward copy of values between grid patches. Restriction is the
injection of data from a fine grid into a coarse grid in regions where fine
and coarse grids overlap. In our vertex centered mesh refinement code, coarse
points coincide with fine grid points and restriction is just a copy operation
of data between grid patches on different refinement levels. Finally,
prolongation refers to the interpolation of data from coarse grids into the
outer boundaries of fine grids to provide boundary data for the \weno{}
reconstruction in the outermost grid points. Since fine grid points are more
densely spaced than coarse grid points, this operation requires
interpolation of values for which we use a simple linear interpolation
method. Synchronization, restriction, and prolongation are
applied in this order after each Runge-Kutta substep to ensure consistent
data between the mesh-refined grids.

In the current implementation in \SpEC{}, prolongation is not mass
conservative and thus leads to mass nonconservation at the refinement
level boundary. In practice, we find this effect to be very small since the
matter density at the grid boundaries is small and mesh refinement is used
only in the very late stages of the simulation. We have not found any
noticeable increase in rest mass nonconservation (see
Fig.~\ref{fig:rest-mass-conservation}) once we turn on mesh refinement,
since the cores of the \ns{s} stay inside the finest refinement level.

\subsection{Gauge conditions}
\label{sec:gauge-conditions}
We evolve the spacetime metric $g^{(4)}_{\alpha\beta}$ using the generalized
harmonic formulation of~\cite{SpECwebsite,Duez:2008rb,Szilagyi:2009qz} in
which the coordinate $x^\alpha$ satisfies the covariant scalar wave equation
\begin{align}
\nabla^\beta \nabla_\beta x^\alpha &= H^\alpha\mpct{,}
\end{align}
for a freely specifiable gauge source function $H^\alpha$. The initial data
are constructed in a gauge $H^\alpha_{\text{initial}}$ that assumes
quasiequilibrium and the existence of a helical
Killing vector. 
At the beginning of the simulation, 
we use $H^\alpha=\hat{H}^\alpha$, where $\hat{H}^\alpha$ is defined
to be a tensor that agrees with $H^\alpha_{\text{initial}}$ in a frame comoving
with the grid, and is constant in time in this frame. Note that $H^\alpha$ is
not a tensor. We 
smoothly transition
from this initial gauge
to a purely harmonic gauge $H_\alpha
\equiv 0$ using a transition function
\begin{align}
\mathcal{F}(t; t_0, \Delta T) &=
\begin{cases}
1 & t < t_0\mpct{,} \\
\exp\bm(-(\frac{t-t_0}{\Delta T})^4\bm) & t_0 \le t\mpct{.}
\end{cases}
\label{eqn:gauge-transition-function}
\end{align}
For the transition to harmonic gauge, we choose $t_0 = 0$,
$\Delta T = 2 \sqrt{d^3 / (2 M_0)}$, where $d$ is the initial coordinate
separation of
the stars and $M_0$ is the baryonic mass of each star (cf. the Keplerian
period $T$ of circular orbit of radius $d$ around a mass $M_0$:
$T \sim \sqrt{d^3/M_0}$). $\Delta T$ is
approximately two
orbital periods which is slow enough to avoid gauge artifacts in the
numerical waveforms. This differs from what is typically done in \bbh{} and
\bhns{} simulations using \SpEC{}, where the simulation directly transitions
to damped harmonic gauge~\cite{Lindblom2009c,Szilagyi:2009qz} around each of
the \bh{},
\begin{align}
H_\alpha &= \mu_L \ln\frac{\sqrt{g}}{N}
t_\alpha - \mu_S g^{(4)}_{\alpha\beta} N^{-1} N^\beta\mpct{,}
\label{eqn:damped-harmonic-gauge-condition}
\end{align}
where $g$ is the determinant of the spatial $3$-metric $g_{ij}$,
$t_\alpha = -N
  \partial_\alpha t$ is the future directed unit normal to the
constant-$t$ surfaces, $N$ is the lapse function and $N^\alpha$ is the
shift vector, near the \bh{(s)}.  We find
that employing the damped harmonic gauge condition reduced the
simulation speed to $\approx 20\%$ compared to the harmonic gauge, due
to damped harmonic gauge inducing a reduction of the allowed time step
size to $\approx 20\%$ of the value allowed in harmonic gauge. We
therefore delay changing into a fully damped harmonic gauge as long as
possible. On the other hand, a pure harmonic gauge condition can lead
to coordinate singularities due to caustics near \ah{} formation and
we found that a ``mild'' version of the damped harmonic gauge
condition lets us avoid caustics while still achieving good evolution
speeds.

The simulations discussed in this paper stay
in harmonic gauge until $t_0 = 22410\,M_\sun$ (approximately
$2.5\,\mathrm{ms}$ or $520\,M_\sun$ before we find an \ah{}) at which time
we
transition to the mild version of the damped harmonic gauge condition
Eq.~\eqref{eqn:damped-harmonic-gauge-condition}.  For the mildly
damped harmonic gauge, we set $\mu_L = \mu_S = 0.2\,M_\sun/M_{\text{ADM}}$,
with $M_{\text{ADM}}$ denoting the \adm{} mass of the system.  This matches the
value chosen in~\cite{East:2011xa} and imposes stronger constraint damping
on smaller black holes that are harder to resolve. We smoothly transition t
the new
gauge using Eq.~\eqref{eqn:gauge-transition-function} with $t_0 =
22410\,M_\sun$, $\Delta T = 200\,M_\sun$.

Finally, just before we expect the \ah{} to form, at $t =
22890\,M_\sun$ ($t-t_{\text{horizon}} \approx -0.24\,\mathrm{ms}$) we add a
fully damped
harmonic instance of Eq.~\eqref{eqn:damped-harmonic-gauge-condition} with
$\mu_L = \mu_S = \left[\ln(\sfrac{\sqrt{g}}{N})\right]^2$ to the
already active mild damped harmonic gauge source. This gauge change is very
rapid with $\Delta T = 30\,M_\sun$.

The complete gauge source term at the time of \ah{} formation is thus
\begin{align}
\label{eq:CompleteGaugeTerm}
H_\alpha &= (1-\mathcal{F}(t; 22410, 200)) H_\alpha^{\text{mild}} + {}
            \nonumber \\
         &\qquad(1-\mathcal{F}(t; 22890, 30)) H_\alpha^{\text{full}}\mpct{,} \\
H_\alpha^{\text{mild}} &= \frac{0.2\,M_\sun}{M_{\text{ADM}}} \left(
    \ln\frac{\sqrt{g}}{N} t_\alpha - g^{(4)}_{\alpha\beta} N^{-1} N^\beta
    \right)\mpct{,} \\
H_\alpha^{\text{full}} &= \left(\ln\frac{\sqrt{g}}{N}\right)^2 \left(
    \ln\frac{\sqrt{g}}{N} t_\alpha - g^{(4)}_{\alpha\beta} N^{-1} N^\beta
    \right)\mpct{.}
\end{align}

\subsection{Excision and postcollapse evolution}
\label{sec:excision-and-post-collapse-evolution}

As described in greater detail in Sec.~\ref{sec:numerical_setup},
during the evolution, we use a complex spectral grid setup surrounding each
\ns{} with a set of concentric spherical shells, transitioning to a
set of shared spherical shells in the wave zone. Such a grid setup is
not well adapted to a single \bh{} having formed after collapse of the
merged \ns{s} since it does not allow us to excise the interior of the
\bh{} from the grid.

Therefore soon after we turn on the full damped harmonic gauge, we switch to
a grid setup containing a single filled sphere at the center of the domain
surrounded by spherical shells and begin
searching
for an \ah{} using an iterative fast flow algorithm
based on~\cite{Gundlach1998}. Once we have detected an \ah{} and
followed its evolution through several time steps, we construct a new
spectral grid consisting only of nested spherical shells, with the
innermost boundary of the innermost shell slightly inside the \ah{},
so that the interior of the \bh{} is excised.  We then
interpolate the spacetime variables from the old spectral grid to the new
one and continue the simulation, keeping the finite volume grid unchanged.

The algorithm for transitioning to a new spectral grid with a single
excision boundary is almost the same as described
in~\cite{Scheel2009,Szilagyi:2009qz,Hemberger:2012jz} for treating the
merger and ringdown of a \bbh{} system after a new common
\ah{} forms around the two individual \ah{}s. In particular, the
excision boundary of the spectral grid changes its shape and size
dynamically to conform to the size and shape of the \ah{} and to
ensure that all characteristic fields of the evolution system are outgoing
(into the horizon), so that excision is well posed without a boundary
condition.  

The main difference between \bh{} evolution for \bbh{} ringdowns versus
\nsns{} remnants is the form of the function that maps
grid coordinates $x^{\igrid}$ to the coordinates $x^{\tilde{\imath}}$
in which
the excision boundary distorts to match the shape of the
\ah{}~\cite{Hemberger:2012jz}.  This
map is
\begin{equation}
x^{\tilde{\imath}} = x^{\igrid}\left(1-f_C(\rgrid)\sum_{\ell m}
Y_{\ell m}(\thetagrid,\phigrid)
\lambda_{\ell m}(t)\right)\mpct{,}
\end{equation}
where $Y_{\ell m}$ are spherical harmonics, $\lambda_{\ell m}$ are 
coefficients, $f_C(\rgrid)$ is a prescribed function, and
 $(\rgrid,\thetagrid,\phigrid)$ are
spherical polar coordinates computed in the usual way from $x^{\igrid}$.
For \bbh{} ringdowns, $f_C(\rgrid)$ is chosen as a simple piecewise linear
function that has discontinuous derivatives at spectral subdomain
boundaries~\cite{Hemberger:2012jz}. 
For \nsns{} remnants, $f_C(\rgrid)$ is a Gaussian that is smooth
everywhere so that the
Jacobian of the map is continuous over the finite volume
domain, which overlaps subdomain boundaries of the spectral domain.

On the finite volume grid on which we evolve the fluid variables, we
``mask'' the excised region. Within the excised region, the metric
variables are set to their value for Minkowski spacetime, while the
density is set to its minimum allowed value $\rho_{0, \text{atmosphere}}$.
As we
want to avoid any dependence of the evolution of the system on that
arbitrarily (and unphysical) choice, we also use modified
interpolation stencils when reconstructing the variables at cell faces
close to the excised region, and when interpolating the fluid
variables from the finite volume grid to the pseudospectral grid. For
the reconstruction of the fluid variables on faces, we use the WENO5
algorithm in the bulk of the simulation, whenever the required
five-points stencil is available. We drop to the second-order MC2
algorithm~\cite{MC} when only a three-points stencil is
available. Finally, we simply copy the value from the neighboring cell
center when we do not have enough points to perform the MC2
reconstruction. On the face directly neighboring the excised region,
the left and right fluxes are both set to their value at the nearest
cell center. The metric variables at cell faces are similarly
interpolated from a three-points symmetric stencil, from a two-points symmetric
stencil, or by copying from the only nonexcised cell center,
depending on the number of nonexcised points available around a given
face. Finally, interpolation from the finite volume grid to the
pseudospectral grid is performed using, when possible, a polynomial
fit to a three-points stencil, with the additional constraint that the
interpolation cannot create new extrema (i.e. the interpolated value
is limited by the minimum and maximum values of the function at the
grid points used in the stencil). When we do not have two points
available on each side of the desired interpolation location, we drop
to linear interpolation, or copy from the nearest nonexcised cell
center when extrapolation is required.

\subsection{Wave extraction}\label{sec:waveform-extraction}
We use the \cce{} method described
in~\cite{Bishop1998, Reisswig:2009rx, Babiuc:2010ze, Taylor:2013zia} to
evolve the gravitational waves emitted by the system from a finite radius to
future null infinity \scriplus. Details on the characteristic method and
its use in \SpEC{} can be found in Sec.~II.3.B of~\cite{Taylor:2013zia}. 
We compute the $(2,2)$-mode $\Psi^{2,2}_4$ of the
Newman-Penrose scalar~\cite{Reisswig2009,Newman1962}
at \scriplus decomposed into spin-weighted scalar spherical harmonic modes.
We then use the fixed frequency integration
method of~\cite{Reisswig:2010di} to compute the gravitational wave strain
$h_{2,2}$
from $\Psi^{2,2}_4$ without taking into account possible drift effects
described in~\cite{Boyle2015a}. Since the total drift of the \bh{} at \ah{}
formation and at the end of the simulation is less than $0.1\,M_\sun$ and
$0.2\,M_\sun$, respectively, the effect of drift is expected to be less than
$1\%$ on the dominant $(2,2)$ mode. Details on the extraction setup for our
simulations are
given in Sec.~\ref{sec:gravitational_wave_signal}.

\section{Initial data}
\label{sec:initial_data}
Initial data for this simulation was produced by a new \nsns{} initial data
solver
based on the work of Foucart et al.\ for \bhns{}
systems~\cite{FoucartEtAl:2008}.
As in that work, we start by considering systems in quasi-equilibrium,
where time derivatives vanish in a corotating frame (this neglect of the
small radial velocity will be addressed later).  We take the metric to
be conformally flat and solve for the lapse, shift, and conformal factor
using the extended conformal thin sandwich (XCTS)
equations~\cite{Pfeiffer2003b}.
The matter in the stars is modeled as a cold ($T=0$) perfect fluid with an
irrotational velocity profile, which is a special case of the more generic
framework used in~\cite{Tacik:2015tja}.  The irrotational limit allows a
straightforward
solution for the velocity and is a more realistic approximation than the
corotating limit, as the effective viscosity of \ns{} matter is
insufficient to synchronize the stars' spins with their orbital
frequency~\cite{Kochanek1992,BildstenCutler1992}.

A particular \nsns{} system is specified in terms of the equation of
state of \ns{} matter, the baryon masses of both stars, and their
coordinate separation.  The solver then uses the above assumptions of
quasi-equilibrium and cold irrotational flow to determine the metric
and matter content of the corresponding spacetime.  Since the initial
data problem consists of several coupled nonlinear equations, the
solver takes an iterative approach, with each iteration composed of a
number of substeps (this procedure closely follows Sec. III.C
of~\cite{FoucartEtAl:2008}, which should be consulted for additional
details).

First, given a trial matter distribution, we find an approximate solution to
the elliptic XCTS equations by taking a single step of a nonlinear solver.
By imposing force balance at the centers of the stars, we then adjust the
orbital frequency of the binary.
We also modify the enthalpy of the matter to drive the locations of its
maxima to the specified stellar centers, thus controlling the stars'
separation.
Finally, we approximately solve the elliptic equations imposing irrotational
flow (constrained to preserve the baryon masses of the stars) and feed the
output to the next step of the iterative procedure.
All of these updates are made using a relaxation scheme to aid convergence.

Throughout the solution process, the numerical data are represented on a
spectral grid
composed of hexahedra, cylindrical shells, and spherical shells, and
approximate solutions to the elliptic equations are provided by the
\textsc{spells} framework~\cite{Pfeiffer2003}.
We periodically evaluate the grid and adjust it to better conform to the stars'
surfaces.  By placing subdomain boundaries close to 
the surfaces, the discontinuities there do not strongly affect the spectral
convergence
of the method for the resolutions used in our simulations.

Additionally, we occasionally perturb the centers of the stars
to reduce the \adm{} linear momentum of the system.  During this procedure,
the centers are \emph{not} constrained to be
colinear with the center of revolution, and the separation of the stars may
deviate slightly from the initially specified value.  Separations reported
here are therefore measured from the final solution.

When constructing strictly quasi-equilibrium data, the solver chooses the
orbital angular velocity $\Omega$ by requiring force balance at the centers
of the stars.  Later, when subsequently refining the initial data to reduce
eccentricity,
$\Omega$ is fixed.
By adding an initial radial velocity, we relax the quasicircular
approximation in order to more accurately model inspiral conditions and reduce
the initial eccentricity.  The magnitude of this velocity is chosen by
evolving each trial set of initial data for a short time in order to measure
the eccentricity of the orbits, then adjusting the (fixed) orbital frequency
and radial velocity according to a heuristic procedure based on the work
of~\cite{Pfeiffer-Brown-etal:2007} and repeating until that
eccentricity is below $10^{-3}$. 
Similar approaches were also used
by the authors of~\cite{Kyutoku:2014yba,Dietrich:2015pxa},
who achieved comparable results.

Results from our code closely match those of the \code{Lorene} solver
by~\cite{GourgoulhonEtAl2001a}.  In particular, we can
accurately reproduce the quasi-equilibrium sequences
of~\cite{TaniguchiGourgoulhon2002,TaniguchiGourgoulhon2003}.

For this study, initial data are generated using a polytropic \eos{} of
the form
\begin{eqnarray}
P &=& \kappa \rho_0^\Gamma\mpct{,}\\
\epsilon &=& \frac{1}{\Gamma - 1} \frac{P}{\rho_0}\mpct{,}
\end{eqnarray}
with $\Gamma =2$ and $\kappa = 123.6\,M_\sun^2$.
Both \ns{s} have a baryon
mass of $M_0 = 1.779\,M_\sun$,
corresponding to an isolated \tov{} star with an \adm{} mass of
$M_\infty=1.64\,M_\sun$, a circumferential radius of $R_{\text{areal}} =
15.1\,\mathrm{km}$ ($10.2\,M_\sun$), and a compactness of
$M_\infty/R_\text{areal} = 0.16$.
Because of the large initial coordinate separation of $81\,\mathrm{km}$
($55\,M_\sun$), the
binding energy is small, $E_b =
6.7\cdot10^{-3} M_\sun$ and the total \adm{} mass of the system is
$M_{\text{ADM}} \approx 2\,M_\infty$.
In the binary configuration, the stars extend
to an isotropic coordinate radius of $12\,\mathrm{km}$ ($8.1\,M_\sun$),
and their centers are
separated by a coordinate distance of $81\,\mathrm{km}$ ($54.5\,M_\sun$).
This system has an
orbital frequency of $\Omega / 2\pi = 133\,\mathrm{Hz}$ ($M_{\text{ADM}} \omega =
0.0132$)
and an eccentricity of less
than $9 \times 10^{-4}$.
Because of the total mass of the system exceeding the maximum mass of a
hypermassive star for a $\Gamma=2$ \eos{}, we expect the merged \ns{} to
collapse to a \bh{} very
quickly~\cite{Baiotti:2009gk,faber:12,Baumgarte:1999cq}.

\section{Results}
\label{sec:results}

\subsection{Dynamics and grid setup}
\label{sec:numerical_setup}

\begin{figure*}
\centering%
\includegraphics{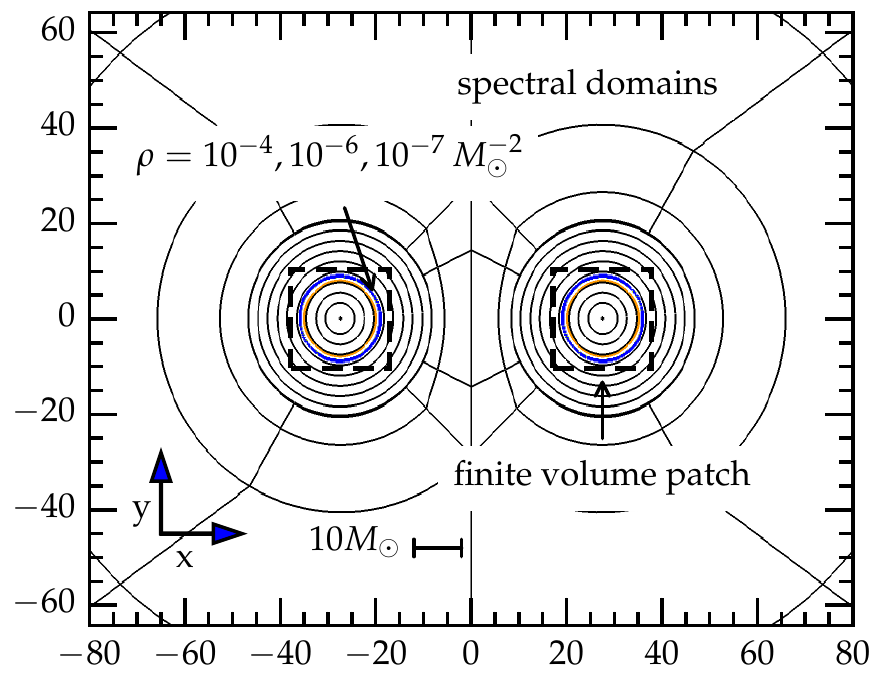}
\includegraphics{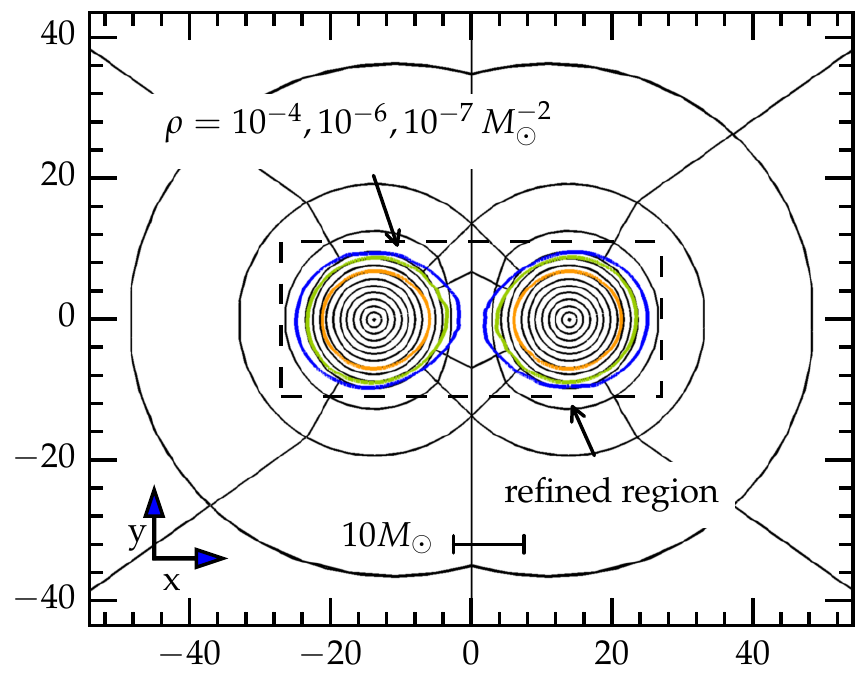}
\includegraphics{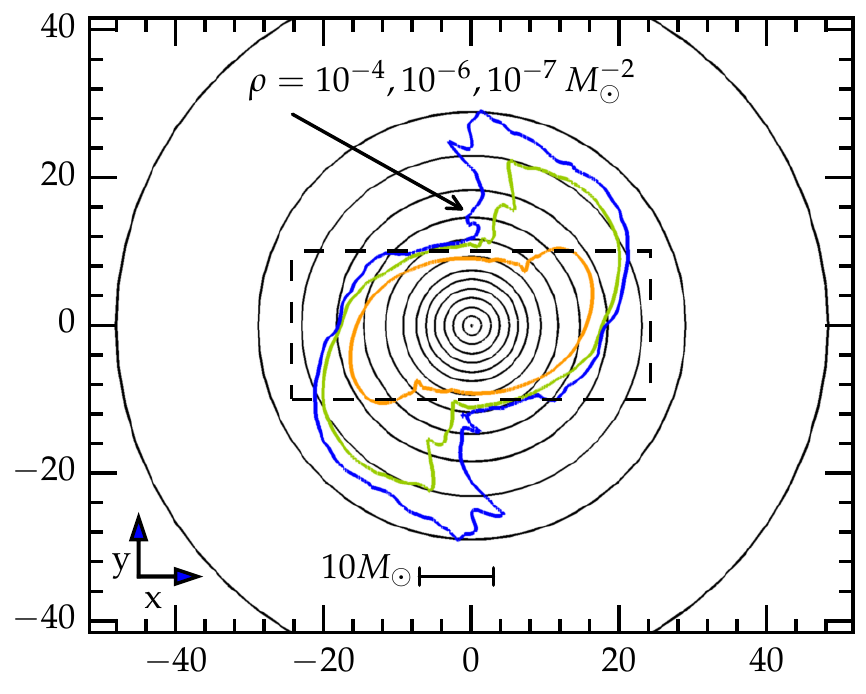} \hfill
\includegraphics{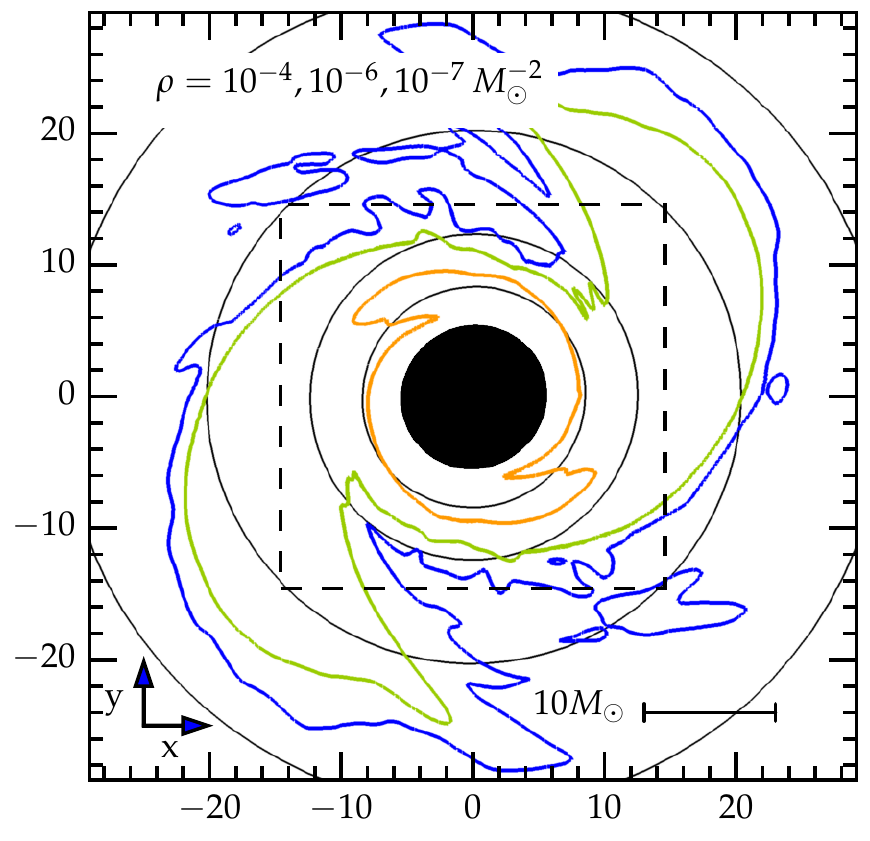}
\caption{
Grid structure for the \Lev0 run in the $z=0$
plane during the simulation. 
We show the spectral grid (solid black lines) and the finite volume grid patches 
that surround the \ns{s}. 
We also show contour lines of the rest mass density $\rho_0$ for
$\rho_0 = 10^{-4},10^{-6},10^{-7}\,(M_\sun)^{-2}$ ($\rho_0 \approx
6\times10^{13},
10^{11},
10^{10}\,\mathrm{g}\,\mathrm{cm}^{-3}$) in orange, green and blue and
a scale bar indicating the size of $10\,M_\sun$ or approximately
$14.8\,\mathrm{km}$.
Top left: Long-dashed boxes 
outline the finite volume grids at the beginning of the simulation. 
Top right: Grid structure after creation of a single finite volume grid
near merger.
Bottom left: Simulation shortly after switching to a single set of nested spherical shells. 
Bottom right: Simulation after the \ah{} has formed. The innermost spherical
shell coincides with the \ah{}.
All but the top left plot show the entire region covered by the coarse
(outer) finite volume mesh, with the fine (inner) mesh outlined using dashed
lines.  }
\label{fig:grid-structure}
\end{figure*}

The evolution of the \nsns{} system proceeds through a series of
stages, each of which is reflected by specific settings used by the
simulations during this phase. Starting from large separation, the
\nsns{} are initially in the inspiral phase characterized by
quasicircular motion around each other.  During this phase, the
separation changes very slowly compared to orbital time scales.
Eventually, the \ns{s} approach each other and come into
contact, which leads to the development of a shear layer in the
contact region and eventually a single merged object. Finally, the
central core of the merged object collapses and forms a \bh{}, which
accretes the remaining material and eventually settles down to a
stationary Kerr \bh{}.  Figure~\ref{fig:grid-structure} and
Table~\ref{tab:grid-structure-stages} show the settings and grid
structures used during the different phases.

For the spectral grid during the inspiral phase, we use an adapted
version of the two-spheres domain used in~\cite{FoucartEtAl:2011}. We
perform simulations using three different resolution levels: \Lev0,
\Lev1, \Lev2. During grid setup and during evolution, we use the
spectral mesh refinement method of~\cite{Szilagyi:2014fna},
decreasing the allowed
truncation error in the spectral expansion of the solution as $e^{-k}$
for resolution level $k$. Thus, the actual number of collocation points
used differs from subdomain to subdomain and is based on features of
the matter distribution and metric variables inside each subdomain.
In contrast to~\cite{FoucartEtAl:2011}, we replace the half of the
grid that covers the \bh{} in~\cite{FoucartEtAl:2011} by a second copy
of the grid covering the \ns{}. Thus, for resolution level \Lev{k},
our domain consists of two filled spheres covering the center of each
neutron star using spherical harmonic basis functions
in the angular
directions,
while the radial dependence is decomposed into one-sided Jacobi polynomials~\cite{Matsushima-Marcus:1995}.

Each filled sphere is
surrounded by eight spherical shells of similar angular and radial
resolution as the inner sphere. Initially the surface of the star is located
in the third shell. 
The far field region around the binary is covered by 20 spherical shells
starting at 1.5 times the initial separation of the stars to 40 times the
initial separation, or approximately $2200\,M_{\sun}$. These shells have
slightly lower angular and radial resolution than the spherical shells
around the \ns{s}.
The region between the innermost shell and the stars is covered by a set of
deformed
cylindrical shells and filled cylinders interpolating between the spheres.
There are a total of $48$ subdomains in the initial setup.
During the simulation, we measure the truncation error in each subdomain
and adjust the subdomain structure by adding and removing points as well as
splitting and joining subdomains such that the measured truncation error is
close to the requested accuracy. Because of the presence of junk radiation at
the beginning of the simulation we keep the grid structure in the outer
spherical shells fixed for one light crossing time of the simulation domain
to avoid the junk radiation triggering mesh refinement and leading to very
high resolution when attempting to resolve the junk radiation.

The finite volume grid during the inspiral phase consists of two
halved cubes with $48, 61, 77$ grid points per half-length of the cube
for the three resolution levels \Lev0, \Lev1, \Lev2. Initially the cube's
sides
are approximately $1.25$ times the diameter of the stars.  This
corresponds to approximately $30, 38, 48$ points across the radius of
the \ns{} and a linear resolution of $326\,\mathrm{m}$,
$252\,\mathrm{m}$, and $192\,\mathrm{m}$ for resolution level \Lev0,
\Lev1, and \Lev2, respectively.  We take advantage of the reflection
symmetry across the $z=0$ plane present in the system to evolve only
the $z>0$ half-space. Function values in the $z<0$ half-space are
computed using the symmetry condition when needed. The region outside
of each \ns{} but covered by the finite volume grid is filled with a
low density atmosphere with rest mass density
$\rho_{0,\text{atmosphere}} = 10^{-13}\,M_\sun^{-2} \approx 10^{-10}
\rho_{0,\text{central}}$, as is common in grid based simulations of
\ns{s}.  For the majority of the simulation the spectral grid structure
consists of sets of spherical shells around each of the \ns{}, and the
finite volume grid consists of one individual grid patch around each \ns{}
as shown in the upper left pane of Fig.~\ref{fig:grid-structure}.

During the simulation we monitor the dephasing in the orbital phase
between resolution levels \Lev0, \Lev1, \Lev2 and interpolate onto a
higher resolution grid once the phase difference increases too
rapidly. For the set of simulations presented in this paper, this
increase in resolution occurs at $t\approx 1.5\times10^4\,M_\sun$
($\approx 38\,\mathrm{ms}$ before the horizon is formed).  At this
point, we increase the resolution of the \Lev0 run to that of the
\Lev1 run ($252\,\mathrm{m}$), that of the \Lev1 run to \Lev2
($192\,\mathrm{m}$) and finally that of the \Lev2 run to
$144\,\mathrm{m}$, which would be the resolution of a \Lev3 run. We
also adjust the requested truncation error in a similar manner such
that the \Lev0 simulation requests a truncation error corresponding to
the truncation error originally requested by the \Lev1 simulation and
similar for the higher resolution simulations.  Once the \ns{s} are
close enough together so that the individual grids touch, we replace
the two grids by a single rectangular grid that covers both
\ns{s}. This is shown in the top right panel of
Fig.~\ref{fig:grid-structure}.  At this time, we increase the
resolution further such that three resolution
levels \Lev0, \Lev1, and \Lev2 use resolutions of $207\,\mathrm{m}$,
$148\,\mathrm{m}$ and
$115\,\mathrm{m}$.
We surround this inner grid by a coarser grid of twice the size but half the
resolution as described in Sec.~\ref{sec:mesh_refinement}.  The coarse grid
captures ejecta and the disk
left behind after \bh{} formation.
At this point we no longer adjust the domain to contain all matter, instead
we hold the physical size of the finite volume grid constant.

\begin{figure*}
\centering%
\includegraphics{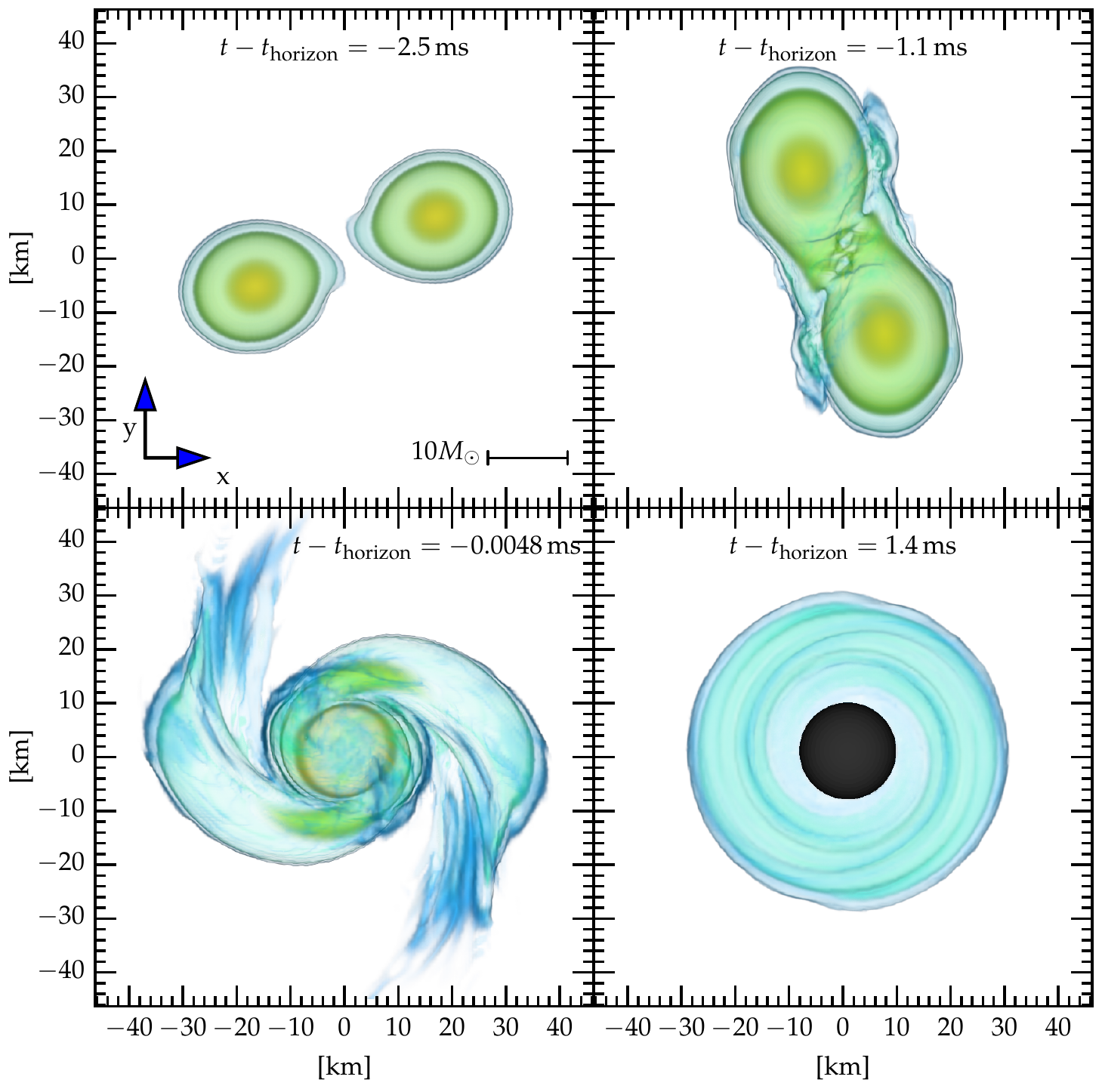}
\caption{Volume rendering of the rest-mass density in the
  late-inspiral, merger, and postmerger phases.  Top left panel:
  Late-inspiral part of the simulation. Shortly before the two \ns{s}
  touch, tidal deformations are clearly visible.  Top right panel:
  Shortly after contact, a characteristic shear layer is formed
  between the two \ns{s}.  Bottom left panel: Shortly before collapse,
  the mass is centered around the origin.  Low-density spiral arms of
  the merger remnant have formed.  Bottom right panel:
  $1.4\,\mathrm{ms}$ after merger the \bh{} has settled down to an
  almost stationary \bh{}.
\label{fig:density-volume-rendering}}
\end{figure*}

When the stars approach each other, they gradually deform,
cf.\ Fig.~\ref{fig:density-volume-rendering} to see the deformation of
the stars.  At some point, the nested spherical shells that are used
in the spectral grid are no longer a good approximation of the stellar
shapes, causing the simulation speed to require more and more
spectral resolution to resolve the deformed stellar
shape. This in turn reduces the allowed time step via the \cfl{}
factor, rapidly reducing simulation speed.  We replace the nested spherical
shells and cylinders around
each star by a single set of concentric spherical shells centered at
the center of the merging binary. During the merger phase, the matter
distribution is very distorted while the metric terms gradually become
centered around the origin. Hence, a spectral grid centered around the
origin deals best with the lack of regularity in the data.  The bottom
left panel of Fig.~\ref{fig:grid-structure} shows the grid layout at
this point.  The 3D density distribution shortly afterwards is shown
in the bottom panel of Fig.~\ref{fig:density-volume-rendering}.

When switching to a single set of spherical shells, we turn off
tracking the orbital separation of the stars and smoothly transition
to a constant scaling factor between the comoving grid coordinates and
inertial coordinates.  The numerical grid is thus no longer
contracting with the stars and the stars move toward each other in the
grid frame. Allowing the stars to move on the grid avoids strong grid
deformation in the region between the stars where a constant volume in
grid space is used to represent the shrinking separation between the
stars while at the same time covering the region far from the stars
with almost constant resolution.  We continue tracking rotation of the
object until an \ah{} is detected at which point we transition to a
coordinate frame that is at rest with respect to an observer at
infinity.

Finally, once the merged object collapses to a \bh{,} we excise the inner
filled sphere and tie the inner boundary to the \ah{} instead (see
Sec.~\ref{sec:excision-and-post-collapse-evolution}). At this point the
setup is identical to what was used in~\cite{FoucartEtAl:2011}.
The bottom right panel of Fig.~\ref{fig:grid-structure} shows the grid structure at this
point.

Table~\ref{tab:grid-structure-stages} lists the computational domains
used in this work.
\begin{table*}
\begin{minipage}{\linewidth}
\centering%
\begin{ruledtabular}
\begin{tabular}{p{4.8cm}ccc}
& Inspiral & Late inspiral & Tidal interaction %
\\
\hline\hline\\[-2ex]
Spectral grid &
2 sets of spheres &
2 sets of spheres &
2 sets of spheres
\\
Finite volume grid &
2 uniform boxes&
2 uniform boxes&
Mesh-refined rectangular box
\\
Control outflows &
Yes&
Yes&
No
\\
Track orbital separation &
Yes&
Yes&
Yes
\\
Finest finite-volume resolution at beginning of segment [$M_\sun$] &
$0.22$, $0.17$, $0.13$&
$0.17$, $0.13$, $0.097$&
$0.14$, $0.10$, N/A\footnotemark
\\
Gauge condition &
Harmonic&
Harmonic&
Harmonic
\\
No. of orbits &
$14$ & %
$8$&   %
${}<1$  \\ %
Orbital angular frequency $M_{\text{ADM}} \omega$ at beginning of segment &
0.014 & %
0.020 &
0.042
\\
\hline\hline
& Plunge & Precollapse & Postcollapse
\\
\hline\hline\\[-2ex]
Spectral grid &
2 sets of spheres &
Spherical shells&
Excised spherical shells
\\
Finite volume grid &
Mesh refined rectangular box&
Mesh refined rectangular box&
Mesh refined square box
\\
Control outflows &
No&
No&
No
\\
Track orbital separation &
Yes&
No&
No
\\
Finite volume resolution at beginning of segment [$M_\sun$] &
$0.14$, $0.099$, $0.078$&
$0.13$, $0.097$, $0.076$&
$0.11$, $0.073$, $0.058$\\
Gauge condition &
Mildly damped harmonic&
Fully damped harmonic&
Fully damped harmonic
\\
No. of orbits &
$1$&   %
${}<1$&  %
N/A \\ %
Orbital angular frequency $M_{\text{ADM}} \omega$ at beginning of segment &
0.045 &
0.092 &
N/A     %
\end{tabular}
\end{ruledtabular}
\footnotetext{We transition to a mildly damped harmonic gauge at fixed
evolution time while transition to a single box is triggered by the finite
volume grids touching. Since the size of the finite volume grids differs
between \Lev{}s, it so happens that the transition to a plunge occurs before
the \ns{s} approach each other close enough to force a single finite volume
box.}
\caption{Stages in the inspiral simulation. For each stage we list the type
of spectral and finite volume grid used, whether we control the amount of
matter flowing off the finite volume grid, whether we monitor the orbital
separation of the \ns{,} the minimum resolution for resolution levels
\Lev0, \Lev1, \Lev2 on the finite volume grid, the gauge condition used,
the
approximate number of orbits the system spends in this phase in the medium
resolution (\Lev1) simulation, and the orbital angular frequency at the
beginning of each segment in the medium resolution (\Lev1) simulation.}
\label{tab:grid-structure-stages}
\end{minipage}
\end{table*}

\subsection{Diagnostics}
\label{sec:diagnostics}
During the course of the simulation we monitor several constraints and
conserved quantities to assess the quality of the simulation.

\subsubsection{Rest mass conservation}
\label{sec:rest-mass-conservation}

\begin{figure}
\centering%
\includegraphics{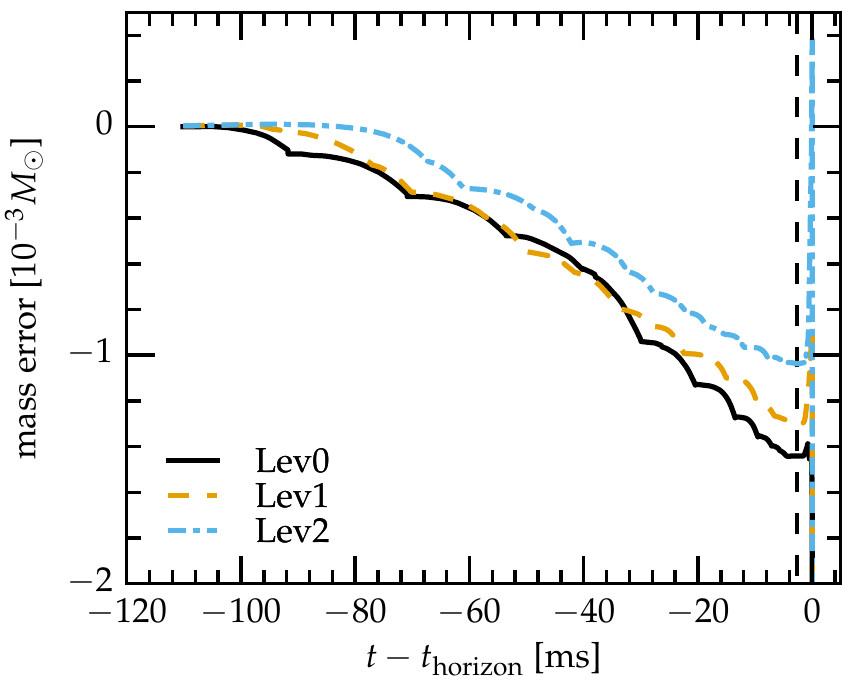}
\caption{Conservation of rest mass for the three resolution levels
  \Lev0, \Lev1, \Lev2.  Mass conservation is very good until the
  horizon forms at which point numerical errors near the center of the
  forming \bh{} lead to spurious mass creation.  The vertical dashed
  line indicates when we switch on the refined meshes. We do not track
  mass conservation after the \ah{} is formed, since material escapes
  from the simulation domain through the \ah{}.  }
\label{fig:rest-mass-conservation}
\end{figure}

We evolve the relativistic rest mass
density $D$ using a conservative scheme~\cite{Duez:2008rb}: \SpEC{} is
therefore expected to exactly conserve total rest mass,
\begin{align}
M_0(t) = \int D \, d^3x\mpct{,}
\end{align}
during the evolution. However, there are several effects that introduce
nonconservative changes to the rest mass density: (i)
we employ (see Sec.~\ref{sec:numerical_setup} for details) a low density
numerical atmosphere of density
$\rho_{0,\text{atmosphere}}$
which can lead to matter creation in the region outside the
\ns{} when the density would drop below $\rho_{0,\text{atmosphere}}$ during
the evolution. We employ an atmosphere density that is sufficiently small
compared to the density at the center of the \ns{s},
$\rho_{0,\text{central}}$, that this effect is expected to be small.
(ii) Matter can reach
grid boundaries and flow off the grid. We employ the remapping procedure
described in Sec.~\ref{sec:comoving_coordinate_system} to control the
amount of matter leaving the system, limiting the matter loss rate through
the boundary to $\dot M_0^{(i)} < 2\times10^{-8}$
through any of the grid boundaries. (iii) The interpolation algorithm
used in the remapping procedure is not mass conservative and
introduces a relative mass change of order $< 10^{-5}$ when
interpolating to a new grid. (iv) Our current mesh refinement
implementation (see Sec.~\ref{sec:mesh_refinement} for details)
uses nonconservative interpolation operators to interpolate data
between the different refined regions. The error introduced by this
interpolation is very small, since the matter density near the
refinement boundaries is very low and does not contribute much to the
total mass.

We find (ii) to be the most important effect during the simulation.
Figure~\ref{fig:rest-mass-conservation} displays the conserved rest
mass over the course of the simulation. We stop monitoring the total
rest mass once an \ah{} is found. At this point rest mass is lost from
the simulated spacetime as matter falls into the \ah{}. The accretion
happens over a short time and matter is rapidly falling into the
\bh{}.  Approximately $2$--$10\,\mathrm{ms}$ ($400$--$2000\,M_\sun$)
after \bh{} formation,
the mass left outside the \bh{} falls below $10^{-3}M_\odot$, which
corresponds roughly to the mass conservation error of our code for
\Lev0 and \Lev1.

\subsubsection{Constraints}
\label{sec:constraint-violations}

In the following section, the $L_2$ volume norm $\|\cdot\|_2$ of a rank
$n$ tensor $T_{\alpha_1\cdots\alpha_n}$ is defined as
\begin{multline}
\|T_{\alpha_1\cdots\alpha_n}\|_2 = \\ \sqrt{
\frac{\int
|
\delta^{\alpha_1\beta_1}\cdots\delta^{\alpha_n\beta_n}
T_{\alpha_1\cdots\alpha_n}
T_{\beta_1\cdots\beta_n}
|^2 \dint^3x}
{\int \dint^3x}
}\mpct{.}
\end{multline}

\begin{figure}
\centering%
 \includegraphics{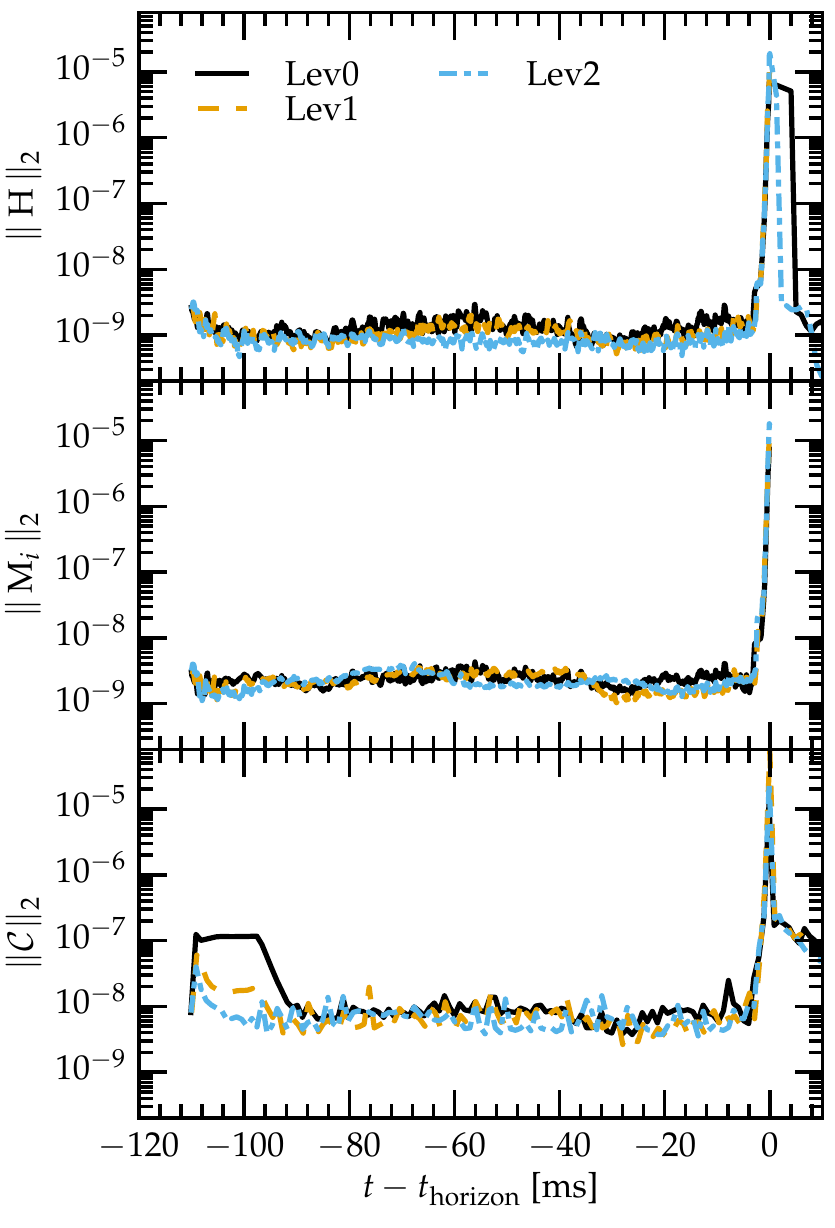}
    \caption{$L_2$ volume norm of the Hamiltonian constraint violation (top panel),
    the square magnitude of the momentum constraint violation (middle panel),
    and the generalized harmonic constraint violation energy (bottom panel).
    The
    origin in time corresponds to \ah{} formation. Clearly visible is the
    increase in constraint violation around this time.
    }
    \label{fig:constraints}      
\end{figure}

\paragraph{\adm{} constraints}
Figure~\ref{fig:constraints} (top panel and center panel) show the
violation of the \adm{} Hamiltonian and momentum constraints, respectively. 
The constraints are evaluated as the $L_2$ norm over the simulation volume. 
The numerical data contain a large amount of noise and we use a moving average
of window width $\Delta t =
1.92\,\mathrm{ms}$ ($40\,M_\sun$) %
to smooth out this high frequency noise.
We observe a spike in the
constraint violations around the time of merger and \bh{}
formation when the spacetime becomes highly dynamic. After excision of the \bh{},
the constraint violations shrink since the inner part of the \bh{} is
no longer part of the numerical or physical domain of dependence and is no
longer included in the computation of the constraint violations.

\paragraph{Generalized harmonic constraint energy}
\label{sec:generalized-harmonic-constraint-energy}

The generalized harmonic formulation of Einstein's equations contains several
constrained quantities. Monitoring these constraints during the simulation
provides us with a useful measure of the faithfulness of our simulations.
Figure~\ref{fig:constraints} (bottom panel) shows the evolution of the $L_2$ norm of
the generalized harmonic
constraint energy as defined in Eq.~(71) of~\cite{Lindblom2006}.

At the beginning of the simulation we see clear convergence and the
constraint decreases for increasing resolution. When \amr{} as
described in Sec.~\ref{sec:mesh_refinement} is activated, i.e.\ after
emission of the junk radiation, the clear convergence is lost, as
constraint violations are no longer dominated by errors in the
spectral domain but instead contain contributions due to matter which
converge much more poorly as resolution increases from \Lev0 to \Lev1
and \Lev2.  Furthermore, as for the ADM constraints, we observe a
spike in the constraint violation around the time of merger and \bh{}
formation when the spacetime becomes highly dynamic. Fortunately, the
constraint violating numerical data are concentrated in the region
that will be interior to the newly formed \bh{}. This is seen as the
sudden drop in the constraint energy once we excise the interior of
the \ah{} from the simulation domain.

\subsubsection{\adm{} integrals}
\label{sec:ADM-mass-integral}

\begin{figure}
\centering%
\includegraphics{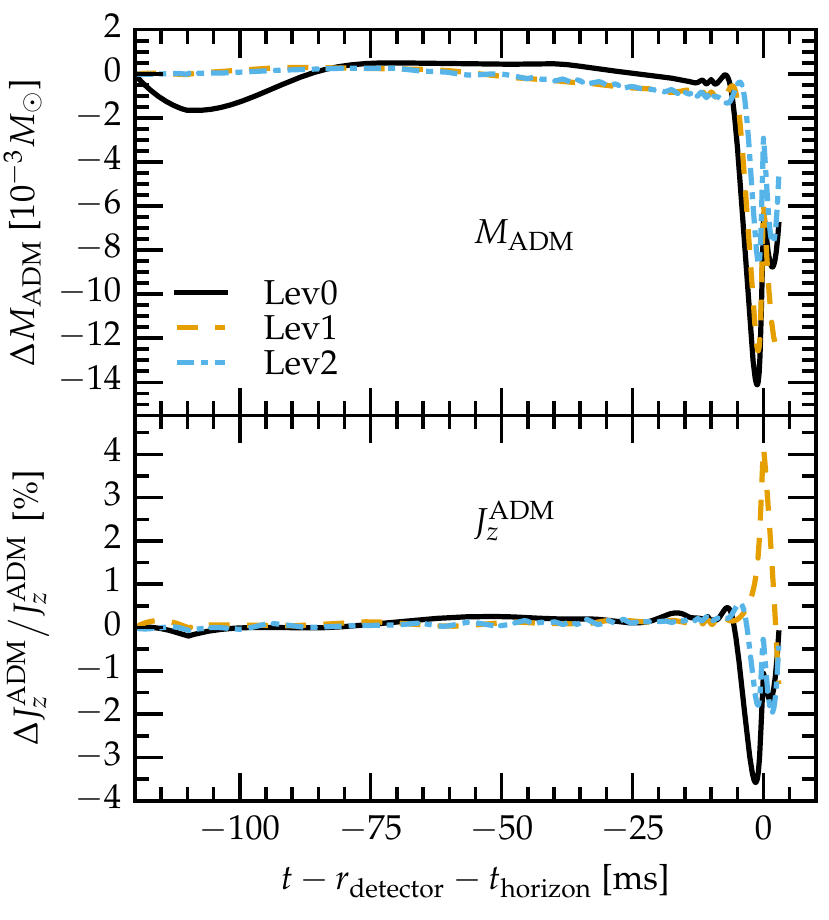}
\caption{Top panel: Mass deficit $M_{\text{ADM}} - \left( M^{\text{NR}} +
M^{\text{rad}} \right)$ computed using a sphere of coordinate
radius $r_{\text{detector}} = 2090\,M_\sun$.
The initial value of $M^{\text{NR}}$ for
simulation \Lev2 is $M^{\text{NR}} = 3.278\,M_\sun$.
The initial decrease in the \Lev0
curve is due to constraints violations in the region $r > 165\,M_{\sun}$,
which are damped away once the spectral
\amr{}~\cite{Szilagyi:2014fna} increases the
resolution in this region. The higher resolution levels \Lev1 and \Lev2 are
of sufficiently high resolution such that this issue does not
arise.
Bottom panel: Relative angular momentum deficit
$1 - \left( J^{\text{NR}}_z + J^{\text{rad}}_z \right) / J^{\text{ADM}}_z$
computed on the same sphere.
Shown are results for the three resolution
levels \Lev0, \Lev1 and \Lev2. The initial value of $J_z^{\text{NR}}$ for
simulation \Lev2 is $J_z^{\text{NR}} = 12.41\,M_\sun^2$ and the total
radiated angular momentum is $J_z^{\text{rad}} = 3.680\,M_\sun^2$.}
\label{fig:ADM-mass-conservation}
\label{fig:ADM-angularmomentum-conservation}
\end{figure}

We also monitor how well the code conserves the total \adm{} mass of the
system during evolution. We approximate \adm{} mass conservation as
conservation of the \adm{} mass surface integrals in the simulation domain
corrected by the radiated energy.
To this end, we evaluate
the \adm{} surface integrals~\cite{Wald84,baumgarteShapiroBook}
\begin{align}
M^{\text{NR}} &=
\frac{1}{16\pi} \oint_{r=r_{\text{ADM}}} \Bigl(
\frac{\partial g_{ik}}{\partial x^j} - \frac{\partial g_{ij}}{\partial x^k}
\Bigr) \delta^{ij} n^k \dint A\mpct{,}
\label{eqn:adm-mass-integral} \\
J^{\text{NR}}_i &=
\frac{1}{8\pi} \epsilon_{ijk} \oint_{r=r_{\text{ADM}}} \bigl(
{K_l}^k - {\delta_l}^k K
\bigr) x^j n^l \dint A\mpct{,}
\label{eqn:adm-angularmomentum-integral}
\end{align}
where $g_{ij}$ is the $3$-metric,
$K_{ij}$ is the extrinsic curvature, $K$ is its trace,
$n^i$ is the outward pointing unit normal vector to the integration
sphere of radius $r_{\text{ADM}} = 2090\,M_\sun$, and $\epsilon_{ijk}$
is the Levi-Civit\'a symbol. We keep track of the radiated
energy and angular momentum in the \gw{} modes up to $\ell = 8$ passing
through the sphere~\cite{Damour:2011fu}:
\begin{align}
M^{\text{rad}} &=
\frac{1}{16\pi} \sum_{\ell,m} \int_0^t \dint t' \left|
\frac{\dint h_{\ell m}(t')}{\dint t'}
\right|^2\mpct{,}
\label{eqn:radiated-energy}\\
J^{\text{rad}}_z &=
\frac{1}{16\pi} \sum_{\ell,m} \int_0^t \dint t' m
\Im \bm{\left(} 
h_{\ell m}(t')
\conj{\left[\frac{\dint h_{\ell m}(t')}{\dint t'}\right]}
\bm{\right)}
\mpct{,}
\label{eqn:radiated-angular-momentum}
\end{align}
where $h_{\ell m}$ is the spin weighted spherical harmonic $(\ell, m)$ mode
of the
gravitational waveform, $\Im(z)$ is the imaginary part of $z$, and $\conj{}$
denotes complex conjugation.
Since we evaluate the \adm{} integral Eq.~\eqref{eqn:adm-mass-integral}
on a surface at a
finite radius, the integrated value depends on time. We correct the value by
the amount of energy radiated through the integration surface and verify
that the sum $M^{\text{NR}} + M^{\text{rad}}$
(and $J_z^{\text{NR}} + J_z^{\text{rad}}$) is constant over time.

The top panel of Fig.~\ref{fig:ADM-mass-conservation} shows
the deficit $M_{\text{ADM}} - \left( M^{\text{NR}} + M^{\text{rad}} \right)$
i.e.\
the failure of the simulation to conserve the \adm{} mass. The lowest
resolution simulation \Lev0 shows an unphysical decrease in the observed
\adm{} mass integral between $-120\,\mathrm{ms} \lesssim t -
t_{\text{horizon}} \lesssim -80\,\mathrm{ms}$ ($-25 \times 10^3\,M_\sun \lesssim t -
t_{\text{horizon}} \lesssim -17 \times 10^3\,M_\sun$),
whose minimum occurs at approximately the same time as the spike in the
generalized harmonic constraint energy.  This decrease is caused by low
resolution in the region $r > 165\,M_{\sun}$ containing the surface used to
evaluate the \adm{} mass integral and vanishes once spectral adaptive mesh
refinement increases the resolution in this region.

The bottom panel of Fig.~\ref{fig:ADM-angularmomentum-conservation}
shows the fractional deficit of the total angular momentum in the $z$
direction, $1 - \left( J^{\text{NR}}_z + J^{\text{rad}}_z \right)
\left(J^{\text{ADM}}_z\right)^{-1}$.  Angular momentum
nonconservation is well below $1\%$ until just before \ah{}
formation. At this point, the increased constraint violations affect
the measurement of the total angular momentum as well.

\subsubsection{Black hole mass and spin}
\label{sec:black-hole-mass-and-spin}

The total mass of the system is well above the maximum mass that can
be supported by postmerger differential rotation and the chosen
\eos{}. A \bh{} forms less than $1\,\mathrm{ms}$ ($200\,M_\sun$) after the
two \ns{}
come into contact and merge.  Within $1\,\mathrm{ms}$ after the
formation of the \ah{}, almost all material has fallen into the
\bh{}. The final mass surrounding the \bh{} is below $10^{-3}\,M_\sun$
for \Lev2.

The final \bh{} settles down to a Kerr \bh{} with Christodoulou mass
of $3.226\pm 0.007 M_\odot$ and spin of $8.743 \pm 0.029 M_\odot^2$,
where uncertainties are estimated as the difference between \Lev2 and
\Lev1.  This corresponds to a dimensionless spin magnitude of
$\chi\approx0.84\pm0.0045$ and is thus well below the extremal Kerr solution.

\subsection{Gravitational wave signal}
\label{sec:gravitational_wave_signal}

\begin{figure*}
    \centering%
    \includegraphics{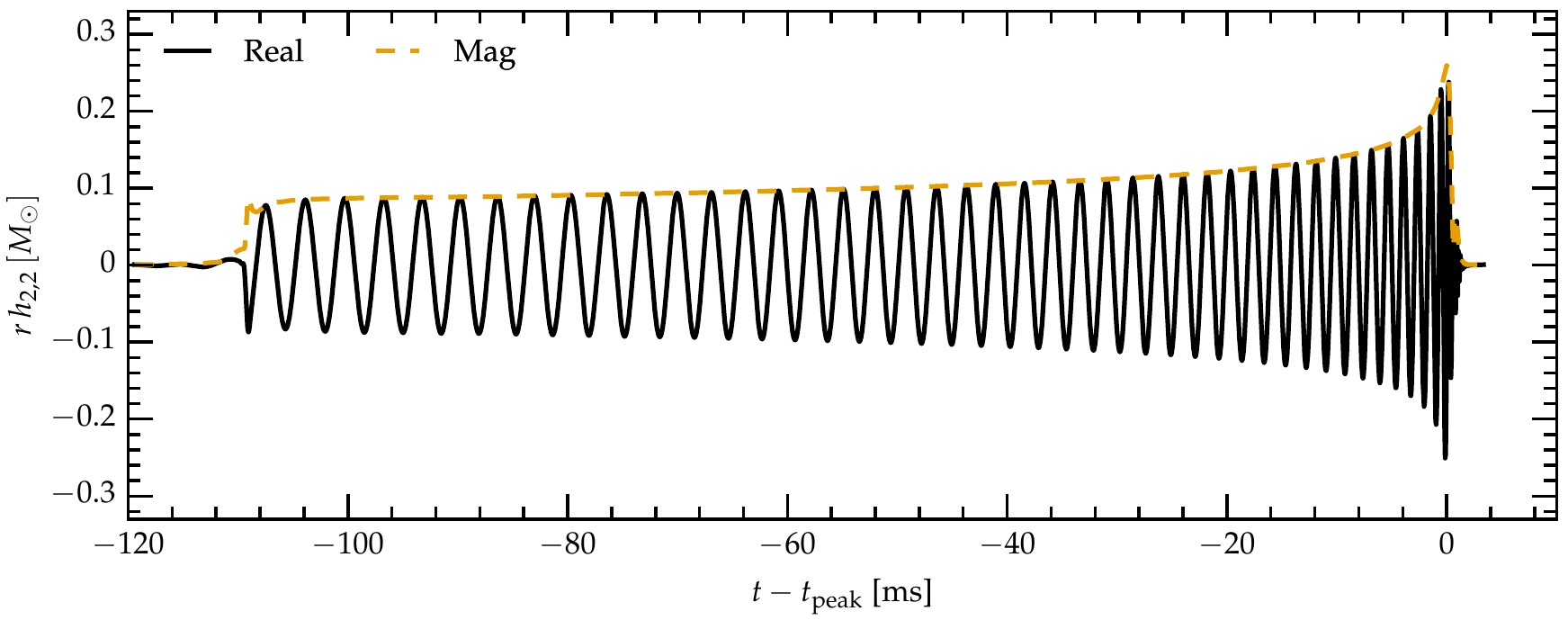}
    \caption{Real part and magnitude of the $(2,2)$-mode of the
    gravitational wave strain $r\,h_{2,2}$ during the inspiral and merger.
    The time of maximum
    amplitude is labeled as $t-t_{\text{peak}}=0$.
    }\label{fig:WaveStrain}
\end{figure*}

Figure~\ref{fig:WaveStrain} displays the $(2,2)$-component of the
spherical harmonic decomposition of the \gw{} strain $h$ at \scriplus
as obtained via the \cce{} method. The waveform lasts for more than 40
GW cycles. We emphasize that this is the longest waveform of a \nsns{}
obtained from a fully general-relativistic simulation.

\begin{figure}
    \centering%
    \includegraphics{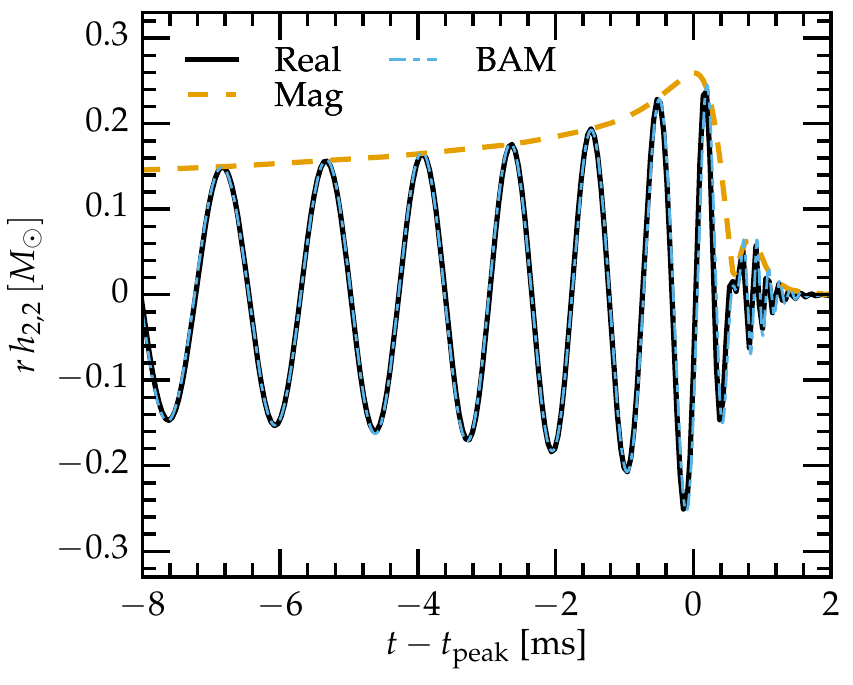}
    \caption{Real part and magnitude of the $(2,2)$-mode of the
      gravitational wave strain $r\,h_{2,2}$ during the late inspiral
      and merger.  The time of maximum amplitude is labeled as
      $t-t_{\text{peak}}=0$. This figure displays a zoom-in of the
      last $\approx 10\,\mathrm{ms}$ of the signal shown in
      Figure~\ref{fig:WaveStrain}, focusing on the final few cycles of
      the inspiral, merger, and ringdown \gw{} signal.  We compare the
      waveform obtained for an identical \nsns{} system
      by~\cite{Bernuzzi:2014owa}. We align in time and phase in the
      interval $-25.8\,\mathrm{ms} \le t \le -13.3\,\mathrm{ms}$,
      minimizing Eq.~\eqref{eqn:rms-phase-difference-integral}.  Both
      waveforms agree reasonably well. During the inspiral the phase
      difference due to the eccentricity of the \bam{} data is around
      $0.1\,\mathrm{rad}$ and even up to merger the phase difference stays
      below $0.25\,\mathrm{rad}$. This is well within the uncertainty of the
      \bam{} waveform of $\pm 0.9\,\mathrm{rad}$.
    }\label{fig:WaveStrainTailEnd}
\end{figure}

In addition to presenting the whole waveform, we also show a zoom-in
around the merger in Fig.~\ref{fig:WaveStrainTailEnd}.  We compare our
waveform with a shorter waveform obtained
by~\cite{Bernuzzi:2014owa,Bernuzzi:2014kca} for the same \nsns{}
system using the finite differencing (spacetime) / finite-volume
(hydrodynamics) \bam{}
code~\cite{Bruegmann2006,Thierfelder:2011yi,Dietrich:2015pxa}.  We
align the waveforms in a time interval $-25.8\,\mathrm{ms} \le t -
t_{\text{peak}} \le
-13.3\,\mathrm{ms}$ ($-5400\,M_\sun \le t - t_{\text{peak}} \le
-2800\,M_\sun$).
This first comparison between the two codes
suffers from the fact that different initial numerical data sets
describing the same physical system were used. Nevertheless, we
observe that after time and phase alignment, the phase difference
stays below $0.25\,\mathrm{rad}$ up to merger.  This is well within
the uncertainty of the \bam{} waveform of $\pm 0.9\,\mathrm{rad}$ and
shows that also around merger, where our error estimate becomes
problematic (see the discussion in Sec.~\ref{sec:convergence_study}),
consistent results can be obtained.

\begin{figure}
\centering%
\includegraphics{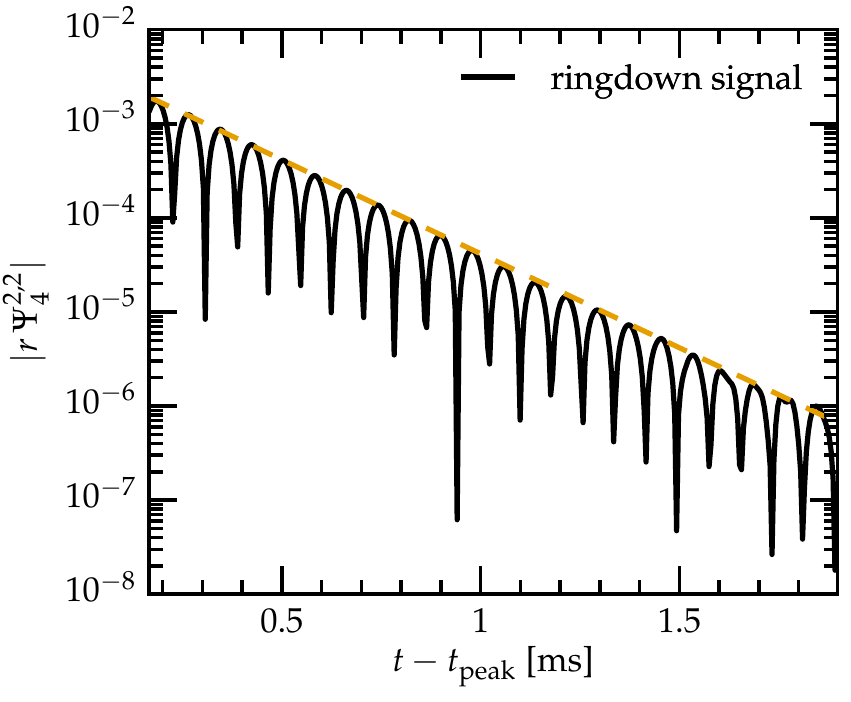}
\caption{Dominant $(2,2)$-mode of the ringdown signal observed in
$\Psi_4$ at $\scriplus$. We observe the ringdown signal for over 3
magnitudes in amplitude before the numerical noise overwhelms the
signal. The dashed line shows the fitted decay behavior
$\exp(-t/\tau_{\text{QNM}})$ with $\tau_{\text{QNM}} = 14.0\,M_{\text{BH}}$.
We find a mode frequency of $M_{\text{BH}} \omega =
0.613$.}\label{fig:QNM22}
\end{figure}

The merged object collapses to a \bh{} within less than
$1\,\mathrm{ms}$ ($200\,M_\sun$) after merger, and the \bh{} then emits a
characteristic
ringdown \gw{} signal.  Our results allow us to estimate the
quasi-normal mode frequencies. We obtain a frequency of $M_{\text{BH}}
\omega = 0.613$ for the $(2,2)$-mode.
This corresponds to within $0.5\,\%$ to the result obtained via \bh{}
perturbation theory and to the value of $0.61454$ stated
in~\cite{Berti2009} for a \bh{} with a dimensionless spin of
$\chi=0.840$. Figure~\ref{fig:QNM22} shows the $(2,2)$-mode of the
ringdown signal in $\Psi^{2,2}_4$ (a similar plot can be obtained for
the \gw{} strain mode $h_{2,2}$). We observe very clean exponential decay
of the dominant mode over more than 3 orders of magnitude of
$\Psi^{2,2}_4$.

\subsection{Convergence}\label{sec:convergence_study}
The physics observable using gravitational wave detectors such as \ligo{}
is primarily encoded
in the phase $\phi$ of a \gw{} signal $h(t) = A(t) \cos
\phi(t)$ and thus the phase accuracy of a simulation is of
primary importance when assessing the quality of a simulation. Since
\gw{} are quadrupolar in nature, for circular orbits, the $(2,2)$-mode
captures the dominant \gw{} signal and its complex phase $\phi$ can be
used to compute (a proxy for) the \gw{} phase $\phi$.

\SpEC{} uses a hybrid spectral -- finite volume scheme that makes it
difficult to assign a unique convergence order to simulations.
\begin{figure}
\centering%
\includegraphics{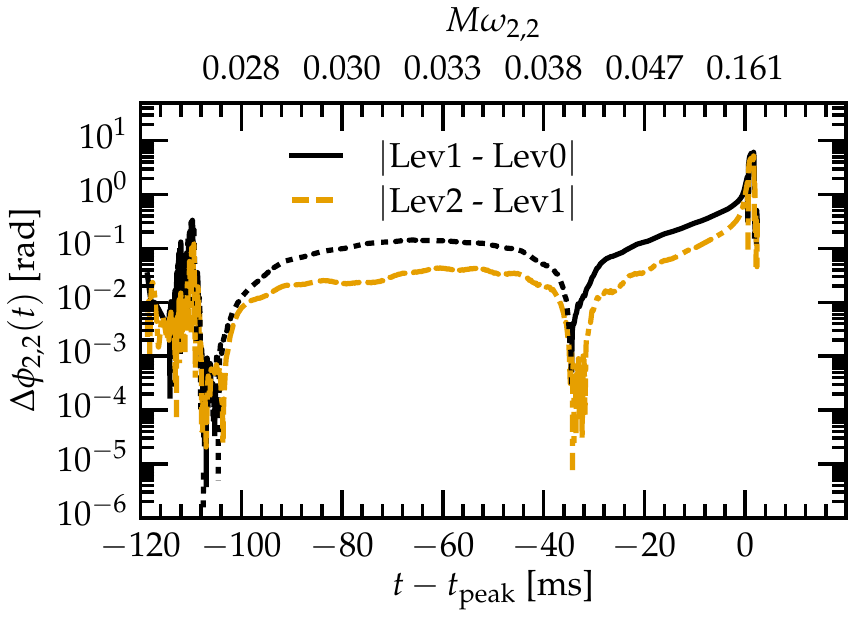}
\caption{ Phase difference $\Delta \phi_{2,2}(t) = \phi_n(t) - \phi_m(t)$
in the $(2,2)$-mode of the \gw{} strain $r\,h_{2,2}$
during the inspiral phase of the simulation among simulations using
resolution level $n$ and $m$.
The solid (black) line shows the phase difference
between the low and medium resolution runs, while the dashed (orange) line
shows the phase difference between the medium and high resolution
runs. Dotted (black) and dash-dotted (orange) line segments indicate time
intervals during which the phase difference  $\Delta \phi_{n,m}(t)$ is
negative.
The upper $x$ axis is labeled
by the \gw{} frequency $\omega_{2,2}$ of the $(2,2)$-mode of the \gw{}
strain of the highest resolution (\Lev2) run.
We observe convergent behavior in the \gw{} phase. However, no
clear convergence order can be assigned. This is most likely due to interactions
between numerical errors in the finite volume hydrodynamics part and in the
adaptively refined
spectral
metric part of the code.
}\label{fig:PhiVsOmega}
\end{figure}
Figure~\ref{fig:PhiVsOmega} shows the phase difference between different
resolution levels \Lev0, \Lev1, \Lev2 in the $(2,2)$-mode of the \gw{} strain
at future null infinity \scriplus. We expect the polynomial error of the
finite volume scheme to dominate the error budget, and thus model the phase
error at each instant in time using a second order polynomial of the form
\begin{align}
\phi = \phi_0 + a_1 \gridspacing + a_2 \gridspacing^2\mpct{.}
\label{eqn:errormodel}
\end{align}
Here, $\gridspacing$ is a measure of the finite volume resolution.
Equation~\eqref{eqn:errormodel} is able to
capture second order convergence of the code in smooth regions of the
flow and first order convergence across shocks and surfaces.  In this
model, $\phi_0$ is the continuum value of the phase and the term $a_1
\gridspacing + a_2 \gridspacing^2$ is the phase error for a simulation
using finite resolution $\gridspacing$. Obviously the model neglects
higher order error terms and the infinite-resolution extrapolated
value $\phi_0$ of the finite resolution \gw{} phases obtained from it
is only an approximation to the true phase.  At very high resolution
($\gridspacing \ll 1$) we expect to recover second order convergence
away from the shock surfaces, which are of lower dimension than the bulk
domain.
Yet at the resolutions used here, the number of grid
points affected by shocks and surfaces is not negligible compared to
the total number of grid points, and a single monomial model for the
error estimate cannot describe the simulation data. A further
complication arises from the fact that the \gw{} strain at \scriplus{}
is given as a function of Bondi time whose relation to simulation time
is complex and depends on both spatial location and
time~\cite{Babiuc:2010ze,Taylor:2013zia}. This makes the assignment of
a single resolution $\gridspacing$ for each time step
difficult. Instead of attempting to extract a value of $\gridspacing$
as a function of Bondi time, we instead use the fact that \cce{}
introduces negligible error compared to the error in the evolution in
the simulation domain~\cite{Taylor:2013zia}. Ignoring the small \cce{}
error, we employ the phase error of the $(2,2)$-mode of the
Newman-Penrose scalar $\Psi_4$ evaluated on a coordinate sphere of
radius $2090\,M_\sun$ as a proxy for the phase error in the \gw{}
strain at \scriplus{} so that there exists a unique resolution
$\gridspacing(t)$ as a function of simulation time.
We find that the change in resolution to control dephasing during the
inspiral at $t - t_{\text{horizon}} \approx -7.9 \times 10^3\,M_\sun$
($-38\,\mathrm{ms}$) is abrupt and
different among the different \Lev{}s.  The differences introduced
by this change are large enough such that the estimated error
at merger is very large ($>0.1\,\mathrm{rad}$ at $t - t_{\text{horizon}}
\approx 7.9 \times
10^3\,M_\sun$ and multiple radians before an \ah{} is detected) if the
change of resolution is included in the data set. Thus we align the
\Lev0 and \Lev1 waveforms to the \Lev2 waveform in the interval
$t_{\text{min}} - t_{\text{horizon}} = -7.7 \times 10^3\,M_\sun \le
t- t_{\text{horizon}}  \le
-2.9\times10^3\,M_\sun = t_{\text{max}} - t_{\text{horizon}}$
($-37\,\mathrm{ms} <
t-t_{\text{horizon}} < -14\,\mathrm{ms}$), corresponding to five inspiral
wave cycles of \Lev2, minimizing the root-mean-square of the phase
difference,
\begin{align}
\left(
\int_{t_{\text{min}}}^{t_{\text{max}}}
|\phi_{\Lev{N}}(t - \Delta t_N) - \phi_{\Lev2} + \Delta\phi_N|^2 \, \dint t
\right)^{1/2}\mpct{,}
\label{eqn:rms-phase-difference-integral}
\end{align}
by varying $\Delta t_N$ and $\Delta\phi_N$~\cite{MacDonald:2011ne}.

\begin{figure}
\centering%
\includegraphics{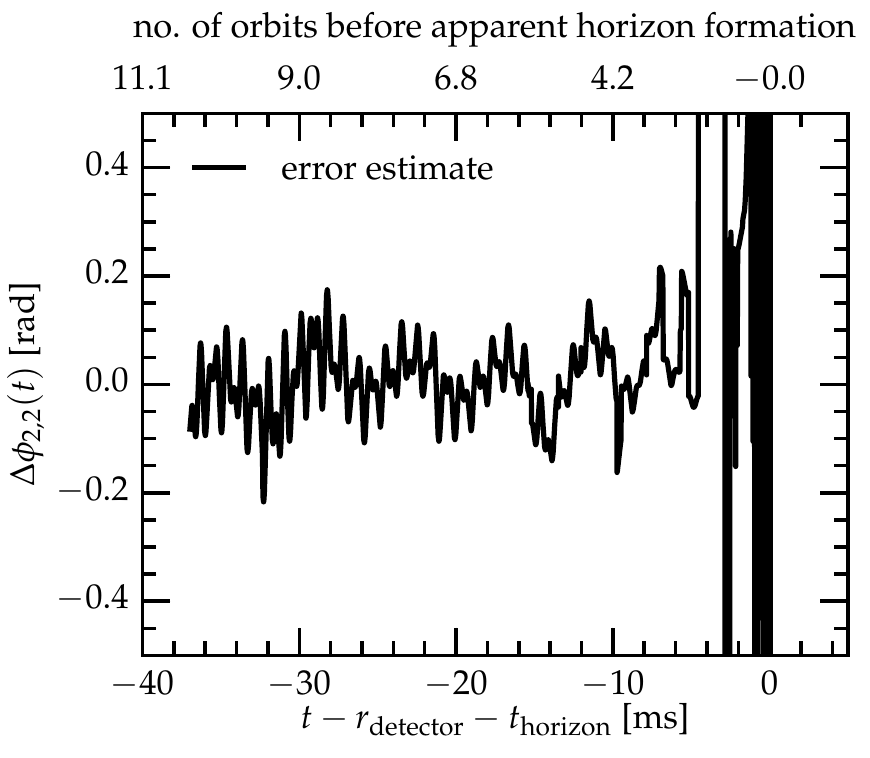}
\caption{Combined error estimate for the phase of $\Psi_4^{2,2}$
after aligning at $t=37\,\mathrm{ms}$ before the \ah{} forms
(see Eq.~\eqref{eqn:rms-phase-difference-integral} for details).
We define $\Delta \phi_{2,2}$ as $\Delta \phi_{2,2}=a_1 \gridspacing + a_2 \gridspacing^2$
according to Eq.~\eqref{eqn:errormodel}. 
$r_{\text{detector}} = 2090\,M_\sun$ is the location of the
extraction surface of the gravitational waves, and retardation is used to
correlate features in the extracted gravitational waves with events in the
strong field
region.}
\label{fig:aligned_error_estimate}
\end{figure}

Figure~\ref{fig:aligned_error_estimate} shows the error estimate
Eq.~\eqref{eqn:errormodel} for the highest resolution run \Lev2 from
the point of alignment onward. Without alignment the estimated phase
error $\Psi_4^{2,2}$ in \Lev2 is significantly larger than
$1\,\mathrm{rad}$. The alignment procedure allows us to estimate the
phase error in a hypothetical simulation that started approximately
$40\,\mathrm{ms}$ ($7800\,M_\sun$) or 11 orbits before \ah{} formation. The
estimated
error is quite small until the last few orbits when approximately
$5\,\mathrm{ms}$ ($1000\,M_\sun$) before \ah{} formation the error estimate
becomes
unreliable.

\begin{figure}
\centering%
\includegraphics{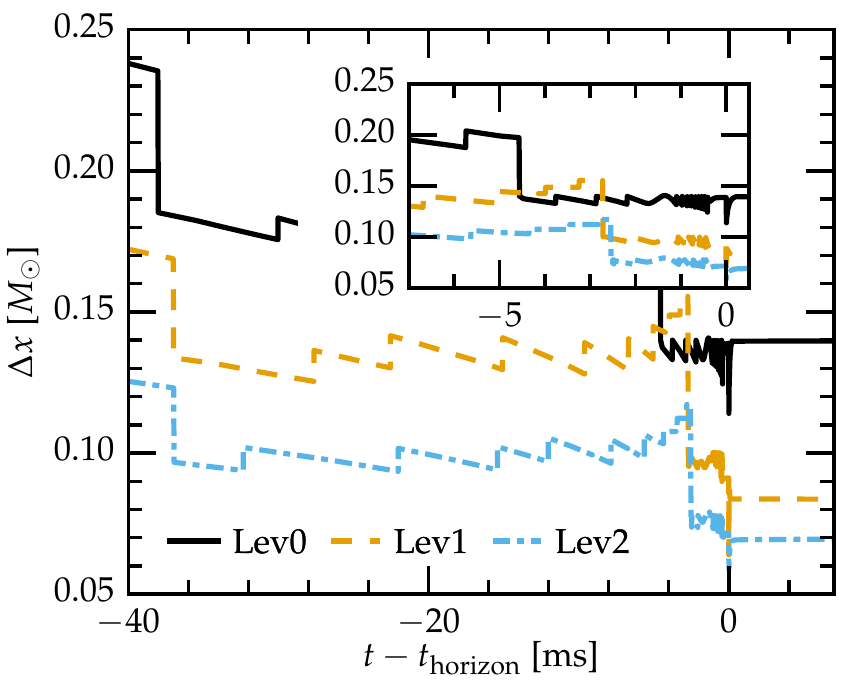}
\caption{Resolution of the finite volume grid covering the \ns{s}
during the
simulation. The inset depicts the resolution in the interval during which we
enable fixed mesh refinement. Refined regions are added based on the
\ns{s'}
separation, and thus the refined region appears first in the \Lev0 simulation
leading to a short time interval during which the ordering of resolutions
is inverted. This leads to problems when attempting to estimate phase errors
in the \gw{} strain if a single monomial dependence of the phase error on
resolution is assumed.}\label{fig:dx}
\end{figure}

The jump in estimated error
coincides with the time we enable fixed mesh refinement on the finite
volume grid, which leads to a situation where temporarily the lowest
resolution run \Lev0 uses a higher resolution than \Lev1.
This is easily visible in
Fig.~\ref{fig:dx} which shows the finite volume resolution during the
final part of the simulations. The inset depicts a zoom-in view of the last
$5\,\mathrm{ms}$ ($1100\,M_\sun$) before \ah{} formation. The slow
increase in resolution over time is due to the inspiral of the \ns{s} and
the jumps are due to remapping of the finite volume grid once material starts
to leave the simulation box. During the period $-5\,\mathrm{ms}\lesssim t -
t_\text{horizon}
\lesssim -2\,\mathrm{ms}$ ($-1100\,\mathrm{ms}\lesssim t -
t_\text{horizon} \lesssim -420\,\mathrm{ms}$) before \ah{} formation, while
\Lev0 is of higher
resolution than \Lev1, the phase evolution between \Lev0, \Lev1, and \Lev2
is also not proceeding as naively expected and a straightforward error
estimate assuming that
$|\Delta \phi_{2,2}(\Lev1, \Lev2)| < |\Delta \phi_{2,2}(\Lev0, \Lev1)|$
yields only an inaccurate estimate for the actual phase error.

\section{Conclusions}
\label{sec:conclusions}
  
In this paper, we presented simulation methods for \nsns{} mergers in
\SpEC{} and discussed the first long \nsns{} inspiral and merger
simulation carried out with \SpEC{}.

The advantages of \SpEC{} compared to other codes are (i) the use of a
hybrid pseudospectral -- finite-volume approach, which reduces
computational costs for the evolution of the spacetime, and, (ii) the
use of comoving coordinates, which eliminates the movement of the
\ns{s} across the numerical domain during the inspiral. 
Currently, \nsns{} simulations
using \SpEC{} are not yet as robust as \bbh{} simulations and require
careful monitoring. This is particularly true for the phase error
whose behavior is not yet fully understood. Further work is required
to compute a robust error estimate for the \gw{} phase.

As an example of \SpEC{'s} capabilities, we presented the longest
\nsns{} inspiral simulation performed to date. Two \ns{s} modeled with
a $\Gamma=2$ \eos{} and a compactness of $0.16$ were evolved for
$\approx 22$ orbits ($44$ wave cycles).  We demonstrated consistency
of our results with shorter, already published results obtained with
the \bam{} code and found remarkable agreement.  A more
detailed study comparing results from multiple different
numerical codes is planned for the future.  Our results show that
\SpEC{} is capable of computing consistent long waveforms for \nsns{}
systems up to and beyond merger.  Such simulations are of great
interest, because long and accurate numerical waveforms are urgently
needed in the new field of \gw{} astronomy. They are essential for
calibrating and validating simpler waveform models employed to detect
\gw{s} and extract information about astrophysics and fundamental
physics from observed \gw{s}.

\acknowledgments We acknowledge helpful discussions with Sebastiano
Bernuzzi, Michael Boyle~\cite{TritonWebsite}, Alessandra Buonanno,
M.~Brett Deaton, Sarah Gossan, Tanja Hinderer, Kenta
Kiuchi, Luis Lehner, Geoffrey Lovelace, Maria Okounkova,
David Radice, Jocelyn Read, Masaru Shibata, Nick Tacik,
and members of our Simulating eXtreme Spacetimes (SXS) collaboration
(\url{http://www.black-holes.org}). This research is partially
supported by NSF Grants No.\ PHY-1068881, No.\ CAREER PHY-1151197, No.\ PHY-1306125,
No.\ PHY-1404569, No.\ PHY-1402916, No.\ AST-1205732, No.\ AST-1333129,
and No.\ AST-1333520,;by the Alfred
P. Sloan Foundation; by the Max-Planck Society; by the Sherman Fairchild
Foundation; and by the International Research Unit of Advanced Future
Studies, Kyoto University.  Support for F.~F. was provided by
NASA through Einstein Postdoctoral Fellowship Grant No.\
PF4-150122 awarded by the Chandra X-ray Center, which is operated by
the Smithsonian Astrophysical Observatory for NASA under Contract No.\
NAS8-03060.  The simulations were performed on the Caltech compute
cluster \emph{Zwicky} (NSF MRI Grant No.\ PHY-0960291), on the
\emph{Datura} cluster of the AEI, on machines of the Louisiana Optical
Network Initiative under Grant No.\ loni\_numrel07, and on \emph{Stampede}
at TACC under NSF XSEDE allocations No.\ TG-PHY990007N and No.\ TG-PHY100033.
All 2D graphs were generated with the \code{Python}-based
\code{matplotlib}~\cite{Hunter:2007} and
\code{ipython}~\cite{PER-GRA:2007} packages.
\code{VisIt}~\cite{Childs:2005ACS,visitweb} was used for 3D and 2D
sliced plots. This paper has been assigned Yukawa Institute for
Theoretical Physics Report No.\ YITP-16-39.

\appendix

\section{Convergence analysis of \tov{} stars}
\label{sec:appendix_TOV}

\begin{figure}
  \centering%
  \includegraphics[width=.495\textwidth]{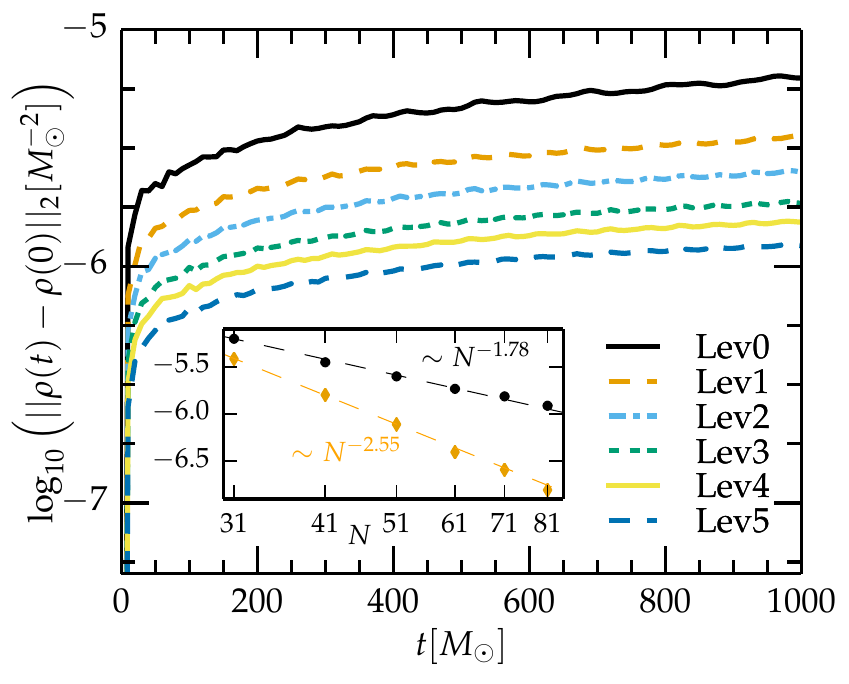}
  \caption{$L_2$ volume norm of the difference between the density profile during the simulation
           and the initial time slice for a stable TOV star for six different 
           resolutions. The inset shows the error computed over the entire
           hydrodynamical
           domain (black circles) and restricted to the inner region of the star 
           with a radius $<5M_\sun$ (orange diamonds) at $t=1000\,M_\sun$
           as a function of the number of grid points used $N$.}
    \label{fig:TOVConvergence} 
\end{figure}

In order to verify our numerical
method and implementation, we present a convergence analysis of an
isolated \tov{} star, reexamining the convergence study
of~\cite{Duez:2008rb}.  Since both spacetime and matter are stationary, any
nontrivial evolution is due to numerical error and in particular the
presence of an atmosphere and sharp surface of the \ns{} influence the
observed evolution. This limits the ability of this test to verify the
expected order of convergence of \SpEC{} as the observed convergence
order depends on the unresolved dynamics at the stellar surface. This fact
is evidenced by finding different convergence orders when including or
excluding the \ns{} surface from the region in which we compute the
convergence order. Nevertheless such a test provides a basic sanity check
for the code and we include it here for this reason.

The initial star is a $\Gamma=2$ polytrope
with a total gravitational mass of $M_\infty = 1.40M_\sun$ and a
radius of $8.1M_\sun$. The star is evolved with a $\Gamma$-law
($\Gamma=2$) \eos{}.  To quantify the numerical error, we compare the
density profile during the evolution with the initial density profile
obtained by solving the \tov{} equation, which can be used as the
exact reference solution.  Our comparison is only meaningful in the
case of the frozen gauge condition described in
Sec.~\ref{sec:appendix_collapse} in which the \tov{} solution is
stationary in the simulation frame. See
Appendix~\ref{sec:appendix_collapse} for a more detailed discussion of the
influence of the gauge. Employing this gauge, the density profile
should stay close to the initial configuration.  In particular, we
compute $\log_{10} \| \rho(t)-\rho(t=0) \|_2 $, where the $L_2$-norm
is either computed within the entire domain or computed inside the
central region of the star, which we define here as $R < 5 M_\sun$.
Figure~\ref{fig:TOVConvergence} summarizes our results.  The main plot
shows the time evolution of the error $\log_{10} \| \rho(t)-\rho(t=0)
\|_2$ computed over the entire star.  The overall error clearly
decreases with increasing resolution.
 
The inset of Fig.~\ref{fig:TOVConvergence} shows errors at
$t=1000M_\sun$ for the entire domain (black circles) and inner region
(orange diamonds) as a function of the number of grid points used.
Theoretically, our finite volume method is limited
to second order because of the choice of using flux values at
face centers for the averaged fluxes when evaluating the right-hand-side
values for the time stepper as well as not distinguishing between averaged
values and reconstructed values when computing the primitive variables from
the conserved ones.
Because of the hybrid grid
approach and discontinuities at the surface of the star, a single
expected convergence order is difficult to define. We observe that,
triggered by the artificial atmosphere, the density profile already
deviates at early times $t\approx50M_\sun$ from the exact solution.
Integrating the error
over the entire star, we observe a convergence order of $\approx 1.75$
(black dashed line).  In fact we cannot expect to obtain high order
or  spectral convergence near the surface of the \ns{}, since the
hydrodynamical variables are not differentiable.  Restricting the
convergence computation to the inner region of the star, we observe a
convergence order around $\approx 2.55$.  We have verified that the
observed convergence order does not change when the atmosphere density is
varied by an order of magnitude.
There is no obvious reason for observing a convergence order higher
than $2$. A possible reason is the fact that the \emph{spatial}
\weno{} reconstruction is of higher order than the overall scheme.
For a system that is
stationary and is very smooth in some regions, the error in the spatially
integrated fluxes and the error incurred during inversion (which are both
only second order convergent) is small
compared to the error incurred during reconstruction of cell averaged
data to cell boundaries (which is fifth order convergent).
Depending  on which error dominates the error budget any convergence order
between $2$ and
$5$ is possible.  Overconvergence is also typically observed
if the resolution is not yet in the convergent regime. However, in
Fig.~\ref{fig:TOVConvergence}, we observe a convergence order of
$2.55$ over a large range of resolutions, which makes this
explanation less likely.  Our analysis shows that when considering the
whole domain, the dominant error comes from the surface of the \ns{},
but the error stays localized and does not spoil the convergence in
the inner region of the star.

\section{Stellar collapse to a black hole in the generalized harmonic formulation}
\label{sec:appendix_collapse}

In this appendix, we address the general problem of simulating the collapse of
a single \ns{} to a \bh{} in the generalized harmonic formulation used by
SpEC as investigated in~\cite{Kaplan:2014wtf}.
The methods presented here are a prerequisite for following the postmerger
evolution to \bh{} formation and ringdown.
We demonstrate convergence and accuracy for
nonrotating, and uniformly rotating test cases.

\subsection{Initial conditions}
\label{sec:appendix_collapse:initial}

We evolve three cases chosen from Baiotti et al.~\cite{Baiotti:2004wn,Baiotti:2008ra}:
a \tov{} case, and two uniformly rotating cases.
The initial stars are modeled by a $\Gamma=2$ polytrope 
and evolved with a $\Gamma$-law ($\Gamma=2$) \eos{}.  Rotating
equilibria are generated using the code
of~\cite{cook92,cook94a}. The parameters
specifying the cases are listed in Table~\ref{tab:sscModels}.

\begin{table}
\centering%
\begin{ruledtabular}
\begin{tabular}{cccccccc}
Case & $M_{\infty}$ & $M_0$  & $\rho_{0,\text{central}}$ & $R_{iso}$  & $r_{p/e}$  \\
\hline\hline
D0  &  1.636  &  1.770  &  3.325$\times 10^{-3}$  &  7.54  & 1.0\\
D2  &  1.728  &  1.913  &  3.189$\times 10^{-3}$   &  8.21   & 0.85 \\
D4  &  1.861  &  2.059 &  3.116$\times 10^{-3}$   &  9.65   & 0.65 \\
\end{tabular}
\end{ruledtabular}
\caption{\label{tab:models} Cases evolved in this study.  All units
  are given in terms of solar masses. $M_{\infty}$ is the ADM
  (gravitational mass),
  $M_0$ is the baryonic mass, $\rho_{0,\text{central}}$ is the maximum
  baryon density of the configuration, $R_{\text{iso}}$ is the coordinate
  radius in isotropic coordinates,
  and $r_{p/e}$  is the ratio of polar to
  equatorial coordinate radii.}
\label{tab:sscModels}
\end{table}

Truncation error alone will cause an unstable stellar equilibrium to evolve
either to a stable equilibrium state or to collapse. In order to demonstrate
convergence, we prefer to induce a resolved evolution toward gravitational
collapse to a \bh{}.  We therefore deplete the fluid pressure by a constant
factor $f_d$.  A pressure depletion can be thought of as a change in the
one-parameter \eos{} $P(\rho_0)$ used to construct the equilibrium. 

In order to avoid violating the constraint equations at the
initial time, this must be done carefully.  Fortunately, the Hamiltonian
and momentum constraints depend on the matter distribution only through
the conserved variables $E$ and $S_i$, defined in
Sec.~\ref{sec:two_domain_approach_to_general_relativistic_hydrodynamics}.  If
the primitive
variables are adjusted but $E$ and $S_i$ are unchanged, the constraints
will be unaffected.  For a rotating star, there are two constraint source
variables ($E$ and $S_{\phi}$ or $S^2$) and two independent fluid variables
(density and rotational velocity).  This, suggests
the following recipe:
\begin{enumerate}
\setlength\itemsep{0pt}%
\item Construct a constraint-satisfying equilibrium for the \eos{}
$P=P_0(\rho_0)$, $\enth=\enth_0(\rho_0)$.
\item Take for one's actual \eos{} $P=f_d P_0(\rho_0)$,
$\enth=1 + f_d [\enth_0(\rho)-1]$.  Since this is the \eos{}
actually used, one may prefer to think that step 1 uses a pressure-enhanced
\eos{}.
\item At each point, re-solve for $\rho_0$ so that $E$ and $S_i$ are the same
as before.
\end{enumerate}
For a perfect fluid
\begin{eqnarray}
E/\sqrt{g} &=& \rho_0 h^e W^2 - P\mpct{,} \\
S_i/\sqrt{g} &=& h^e W \rho_0 u_i\mpct{,}
\end{eqnarray}
where $W$ is the Lorentz factor.

One eliminates $W$ using
\begin{equation}
\label{PressureDepletionNewW}
  W^2 = \frac{E/\sqrt{g}+P}{\rho_0 \enth}\mpct{.}
\end{equation}

The new density is obtained by solving for the root of the one-dimensional equation
\begin{equation}
\label{PressureDepletionNewRho}
S^2 = (E + \sqrt{g}P)[E + \sqrt{g}(P-\rho \enth)]\mpct{.}
\end{equation}

Hence, one solves for $\rho_0$ using Eq.~\eqref{PressureDepletionNewRho}
and uses this to find the new rotation rate via Eq.~\eqref{PressureDepletionNewW}.

\subsection{Gauge conditions and dynamics}

We investigated a series of five different gauge conditions in order to
study the coordinate dynamics during gravitational collapse and attempt
to determine what condition leads to the most robust simulation of
\bh{} formation.  All simulations start from the same gauge, set by
the initial conditions, but we have the option of transitioning to another
gauge, as described in Sec.~\ref{sec:gauge-conditions}, during the
simulation.
The gauge conditions are denoted
\begin{description}
\item[Frozen] for a frozen gauge $H_\alpha(t) = H_\alpha^{\text{initial}}$;
\item[Harm] for transition to a pure harmonic gauge $H_\alpha = 0$;
\item[Full] for transition to the damped harmonic gauge given by
Eq.~\eqref{eqn:damped-harmonic-gauge-condition};
\item[Shift,] which transitions to a gauge where only the spatial components of
Eq.~\eqref{eqn:damped-harmonic-gauge-condition} are imposed;
\item[Slice,] which only imposes the damped harmonic condition on the $t$
component of Eq.~\eqref{eqn:damped-harmonic-gauge-condition}.
\end{description}
In all cases but the frozen gauge, we transition away from
the initial gauge using the roll-off function Eq.~\eqref{eqn:gauge-transition-function}
choosing a value
for $\Delta T = 10.0\,M_\sun$. This results in the $H_\alpha^{\text{initial}}$
contribution being driven to zero within roundoff precision by $t = 30.0\,M_\sun$.
For the shift only, slicing only, and fully damped harmonic gauge conditions, we 
transition to (``roll on'') the new gauge with $\Delta T =  25.0\,M_\sun$,
which is about half of the time to \bh{} formation,
so that our damped harmonic gauge condition has fully ``kicked in'' by the time
of \bh{} formation.

\begin{figure}
  \centering%
  \includegraphics{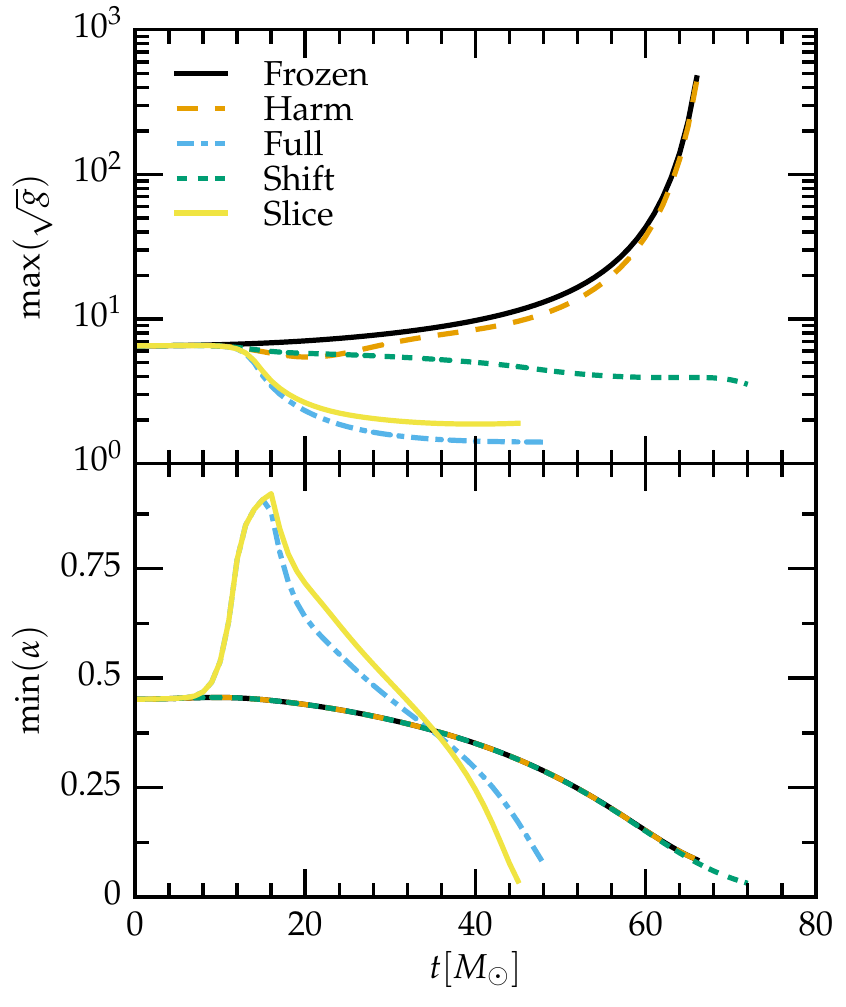}
  \caption{Evolution of the (top panel) maximum of the three-volume element,
           $\sqrt{g}$, and (bottom panel)
           minimum of the lapse, $\alpha$, for the \tov{} case D0.
    \label{fig:sscMetricVars}}
\end{figure}

We will begin our discussion with results for D0, and since all
cases display a similar behavior, we only point out differences
between cases where appropriate.

The first observation is that
the evolution of the central baryon density during the collapse simulations
before \ah{} formation proceeds differently for the
different gauge choices.  Although the proper time
for the central density to reach any value is gauge independent, the
gauge choice affects the evolution of the lapse $\alpha$ as seen in
Fig.~\ref{fig:sscMetricVars}, and
therefore the central density growth curve as a function of coordinate time.
This observation shows that we can expect different gauge choices to affect
the dynamics of \ah{} formation since more rapid evolution %
is harder for the code to resolve when the flow of time,
given by the lapse, differs strongly between different regions of the
simulation domain. This eventually leads to steep spatial gradients as
different fluid regions evolve apart from each other.

Figure~\ref{fig:sscMetricVars} summarizes the dynamics
of case D0 as a function of gauge choice.
The top panel shows the maximum of the spatial
volume element, $\sqrt{g}$. This quantity determines how much physical
volume is represented per unit of coordinate volume. Thus the
larger the value of $\sqrt{g}$ is, the lower the effective resolution is,
since a larger amount of physical volume is represented by a unit of
coordinate volume. For a well resolved simulation, $\sqrt{g}$ must not
increase drastically. Otherwise, the coordinate evolution is deresolving
the simulation (effectively the grid is being fatally stretched out and
distorted in physical space). One can see in
Fig.~\ref{fig:sscMetricVars} that this grid stretching is exactly what
happens during collapse in pure harmonic and frozen gauges.  
The damped harmonic gauge is designed
to dynamically damp $\log(\sqrt{g}/\alpha)$ to zero, and thus
drive $\sqrt{g}/\alpha$ to order unity. This can be understood by looking at the evolution of the
lapse function in the  bottom panel of
Fig.~\ref{fig:sscMetricVars}. It is interesting to note that the damped
harmonic shift condition exhibits a lapse evolution similar to
harmonic or frozen gauge, but a distinct evolution for $\sqrt{g}$.  
Imposing the damped harmonic condition on the $t$ component (the Slice gauge choice) leads to 
an evolution of $\max(\sqrt{g})$ and $\min{(\alpha)}$ qualitatively 
similar to that produced by the damped harmonic condition. 
In general, we find that the damping of the coordinate dynamics imposed by the full, shift or slice condition
is enough to prevent the divergence of the volume element
as the \bh{} forms for case D0, and similar (not shown in the paper) for
the rotation cases.

In practice we find that the damped harmonic gauge
leads to the most robust \bh{} formation simulations. Because of this, we
use it in our \nsns{} simulations when the merged object is about to collapse
to a \bh{}, yet we use harmonic gauge during the earlier phase since it yields
faster simulations as described in Sec.~\ref{sec:gauge-conditions}.
For the D0 case, an \ah{} is first found at coordinate time
$t = 48\,M_\sun$ for the evolution in damped harmonic gauge. At coordinate time
$t = 50\,M_\sun$, after the \ah{} has been found successfully a total of 8 times,
the collapse evolution terminates.
At this point, enough information is available to properly excise the
\bh{} and initialize the ringdown simulation. At the time of \ah{}
formation, the constraint violation has increased only by a factor of 
10 in damped harmonic gauge. In contrast,
by the time the constraints have increased by the same factor in the
harmonic and frozen gauge runs, an \ah{} has yet to be found and the code
eventually crashes due to large constraint violations. 
Imposing the damped harmonic condition on the $t$ component (slice choice) or
the spatial component (shift choice) is also sufficient for following black hole formation.
In our simulations, when we impose the damped harmonic condition only on the $t$ component, the \ah{}
forms earlier in terms of coordinate time, but at the time the \ah{} 
is found, the constraint violations have already increased by $1$ order of magnitude
compared to the damped harmonic gauge.
When the damped harmonic shift condition is imposed, the \ah{} is found 
at later times, and the constraints are at the same order as for the damped harmonic gauge. 

Our finding that evolving in damped harmonic gauge is advantageous to resolve
\bh{} formation extends and confirms the results
of Sorkin~\cite{Sorkin:2009wh}, who found that the damped harmonic gauge is
particularly robust when forming \bh{s} from a complex
scalar field in axisymmetric simulations.

\subsection{Convergence of simulations}

\begin{figure}
  \centering%
   \includegraphics{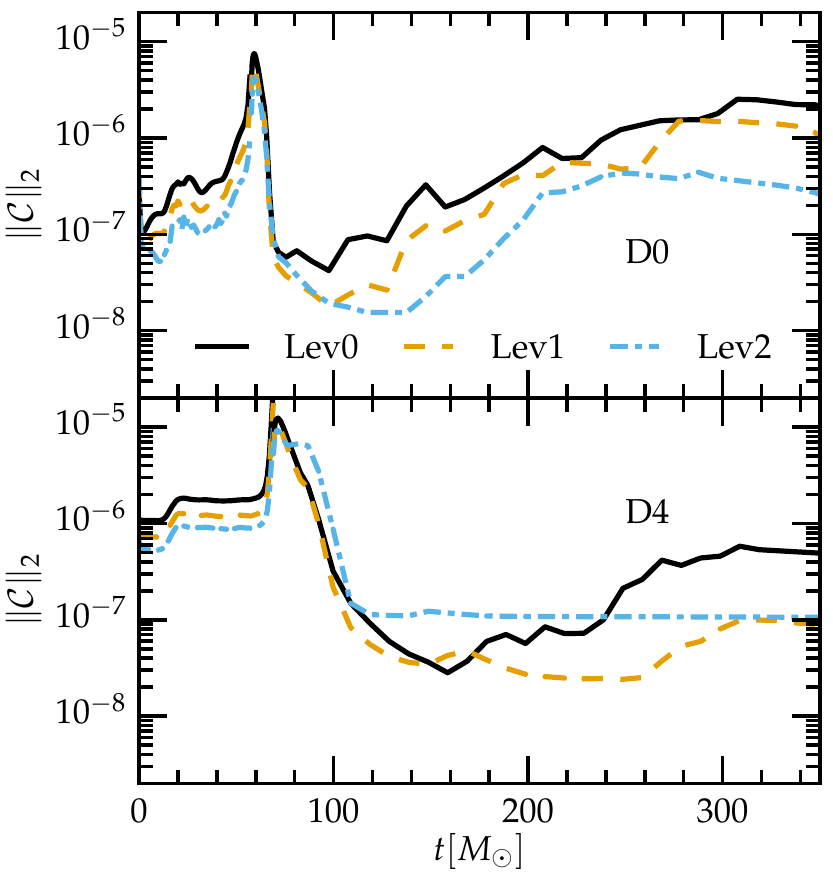}
  \caption{$L_2$ norm of the normalized generalized harmonic
    constraints for the \tov{} case D0 (top) and the
    uniformly rotating case D4 (bottom panel).  Note
    that the maximum in the constraints corresponds to the time of
    \bh{} formation.}
    \label{fig:sscD0D4GhCe}
\end{figure}

To study the convergence properties of \SpEC{} for the single star case
studied here, we conduct three simulations of the same physical setup using
three different resolutions. We use $\Delta x = 300\,\mathrm{m}$,
$250\,\mathrm{m}$, and $200\,\mathrm{m}$ ($\Delta x = 0.20\,M_\sun$,
$0.17\,M_\sun$, and $0.14\,M_\sun$) in the finite volume grid.
In the spectral grid, we use $N_r = 16, 18, 20$ grid points in the radial direction of
each spherical shell and an angular resolution including up to $\ell = 10,
12$, and $14$ spherical harmonic modes for all but the spherically symmetric
case $D0$ for which we do not increase angular resolution with increasing
$\Lev{}$ number. No spectral \amr{} is used for this test before an \ah{}
is found to simplify the
convergence behavior. After an \ah{} forms, we use \amr{} to adjust the
number of grid points in the radial direction $N_r$ but not the spherical
harmonic multipole number $\ell$. This ensures that the region around the
\ah{} is resolved well enough to avoid code simulations failures due to large
numerical errors.
In Fig.~\ref{fig:sscD0D4GhCe} we show plots which
demonstrate the convergence of the simulations with resolution.  
We show the $L_2$ norm of the generalized harmonic constraints for the \tov{}
case D0 and the rotating case D4. Case D2 shows a similar behavior.
The maximum in constraint violation corresponds to the time of
\bh{} formation, after which we excise the interior of the \bh{} from the
numerical domain. This reduces the amount of constraint
violation on the grid. 
Both plots show clear evidence of convergence of the
constraints with an increasing resolution level before
\bh{} formation.
After \bh{} formation, case D0 and similarly
cases D2 and D4 show an overall decrease of constraint violation with
increasing
resolution level, yet the detailed evolution with time varies slightly
between resolution levels. Partially this is due to \amr{} which
occasionally chooses identical resolution for individual subdomains for
different resolution levels. This happens when the estimated truncation
error in the affected subdomains is just above/below the threshold for
derefinement/refinement for two resolution levels. We also observe a pulse
of nonconvergent constraint violation in the outer spherical shells which
eventually leaves the simulation domain, yet contributes to the observed
constraint violation. Case D4 shows a much stronger nonconvergent behavior
for \Lev2 for which we unfortunately are not able to provide a simple
explanation.

\subsection{Gravitational waveforms}

\begin{figure}
  \centering%
   \includegraphics{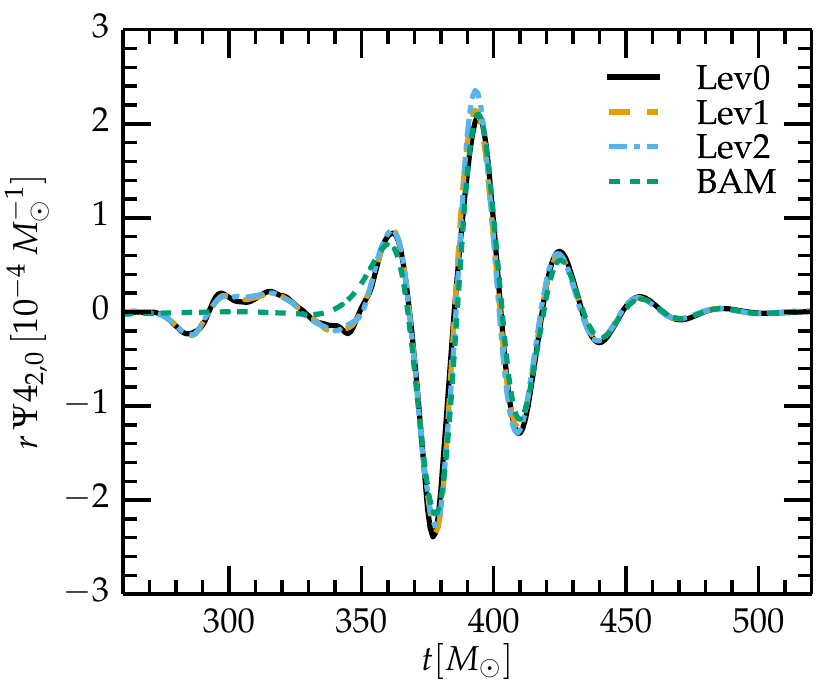}
  \caption{Gravitational wave emitted during the collapse of the uniformly
           rotating \ns{} (case D4). We show different resolutions
           and compare our waveform with the published results
           of~\cite{Dietrich:2014wja}. 
           The \bam{} waveform of~\cite{Dietrich:2014wja} is shifted in time such
           that amplitude maxima coincide.}
    \label{fig:sscD4Psi4}
\end{figure}

Finally we discuss briefly the \gw{s} emitted from the collapse
of our single \ns{} cases.
As pointed out in, e.g.,~\cite{StPi85,Seidel:1987in,Seidel:1988za,Seidel:1990xb} 
the \gw{s} emitted during the collapse of a rotating \ns{} have a particular simple
structure consisting of a precursor-burst-ringdown  pattern. 
We find this characteristic structure in our simulations. See 
Fig.~\ref{fig:sscD4Psi4} for visualization of the $(2,0)$-mode of $\Psi_4$ for the D4 case.
In addition to our results, we present the waveform of~\cite{Dietrich:2014wja}
for the same case. As for the comparison of the \nsns{} waveform
in Sec.~\ref{sec:gravitational_wave_signal},
Ref.~\cite{Dietrich:2014wja} uses the \bam{} code.
The extraction radii are slightly different between the waveforms, 
while our waveforms are extracted at a fixed coordinate radius of $r=259M_\odot$, and the
\bam{} waveform is extracted at $r=250M_\odot$.
However, the main difference is caused by the artificial
pressure perturbation. Here, we perturb the \ns{} according to the 
discussion in Appendix~\ref{sec:appendix_collapse:initial} and set $f_d=0.9$
for a pressure depletion of $10\%$.
In~\cite{Dietrich:2014wja}, the pressure is simply
decreased by $0.5\%$.
Although the ansatz of~\cite{Dietrich:2014wja} does not ensure that the constraint 
equations are satisfied on the initial slice, 
the pressure perturbation is smaller than in our setup. %
This explains differences in the early part of the waveform at times $t\lesssim350M_\odot$. 
In fact, this part of the waveform is unphysical and solely
caused by the perturbation of the rotating \ns{}. 
After $t=350M_\odot$, the \SpEC{} and \bam{} waveforms agree well
and the maximum amplitude difference is $\lesssim 2 \times 10^{-5}$. 

The results presented in these appendixes show that \SpEC{} is well suited to study the collapse
of a \ns{} into a \bh{}. This is of great importance since in most realistic astrophysical scenarios, the 
merger remnant formed after the merger of two \ns{s} will eventually collapse to a \bh{} either on a dynamical or secular timescale, depending on its mass and on the nuclear EOS.

\bibliographystyle{apsrev-edit}

\end{document}